\documentstyle[12pt]{article}
\input epsf.sty
\begin{document}

\thispagestyle{empty}

\renewcommand{\thefootnote}{\fnsymbol{footnote}}

\begin{flushright}
{\small
SLAC--PUB--9949\\
October 2003\\}
\end{flushright}

\vspace{0.5cm}

\begin{center}
{\large \bf  Production of $\pi^+$, $\pi^-$, $K^+$, $K^-$, p and $\bar{\rm p}$
in Light ($uds$), $c$ and $b$ Jets from $Z^0$ Decays\footnote{Work supported by
Department of Energy contracts:
DE-FG02-91ER40676, DE-FG03-91ER40618, DE-FG03-92ER40689, DE-FG03-93ER40788,
DE-FG02-91ER40672, DE-FG02-91ER40677, DE-AC03- 76SF00098, DE-FG02-92ER40715,
DE-FC02-94ER40818, DE-FG03-96ER40969, DE-AC03-76SF00515, DE-FG05-91ER40627,
DE-FG02-95ER40896, DE-FG02-92ER40704;
  National Science Foundation grants:
PHY-91-13428, PHY-89-21320, PHY-92-04239, PHY-95-10439, PHY-88-19316,
PHY-92-03212;
  The UK Particle Physics and Astronomy Research Council;
  The Istituto Nazionale di Fisica Nucleare of Italy;
  The Japan-US Cooperative Research Project on High Energy Physics;
  The Korea Science and Engineering Foundation.}}

\vspace{1.0cm}

 {\bf The SLD Collaboration$^{**}$}\\
Stanford Linear Accelerator Center,
Stanford University, Stanford, CA~94309

\end{center}

\vfill

\begin{center}
{\bf\large   
Abstract }
\end{center}

We present improved measurements of the differential production rates
of stable charged particles in hadronic $Z^0$ decays, and of charged pions,
kaons and protons identified over a wide momentum range using the SLD
Cherenkov Ring Imaging Detector.
In addition to flavor-inclusive $Z^0$ decays, measurements are made for $Z^0$
decays into light ($u$, $d$, $s$), $c$ and $b$ primary flavors, selected using
the upgraded Vertex Detector.
Large differences between the flavors are observed that are qualitatively
consistent with expectations based upon previously measured production and decay
properties of heavy hadrons.
These results are used to test the predictions of QCD in the Modified Leading
Logarithm Approximation, with the ansatz of Local Parton-Hadron Duality,
and the predictions of three models of the hadronization process.
The light-flavor results provide improved tests of these predictions,
as they do not include the contribution of heavy-hadron production and decay;
the heavy-flavor results provide complementary model tests.
In addition we have compared hadron and antihadron production 
in light quark (as opposed to antiquark) jets.
Differences are observed at high momentum for all three charged hadron species,
providing direct probes of leading particle effects, and stringent constraints
on models.

\vfill

\begin{center}

Submitted to {\it Physical Review D}\\

\end{center}

\vfill\eject

\pagestyle{plain}

\section{Introduction}

The production of jets of hadrons from hard partons produced in high energy
collisions is understood qualitatively in terms of three stages.
Considering the process $e^+e^- \rightarrow q\bar{q}$,
the first stage, ``fragmentation", involves the radiation of gluons from the
primary quark and antiquark,
which in turn may radiate gluons or split into $q\bar{q}$ pairs until their
virtuality approaches the hadron mass scale.
This process is in principle calculable in perturbative QCD, and three
approaches have been taken so far:
i) differential cross sections have been calculated \cite{ert} for the
production of up to 4 partons to second order in the strong coupling $\alpha_s$,
and leading order calculations have been performed recently for as many as 6
partons (see e.g. \cite{moretti});
ii) certain parton distributions have been calculated 
to all orders in $\alpha_s$ in the Modified Leading Logarithm
Approximation (MLLA)~\cite{mlla};
iii) ``parton shower" calculations \cite{nlla} have been implemented in Monte
Carlo simulations that consist of an arbitrary number of
$q\!\rightarrow \! qg$,
$g\! \rightarrow \! gg$ and $g\! \rightarrow \! q\bar{q}$ branchings,
with each branching probability determined
from QCD at up to next-to-leading logarithmic order.

In the second stage these partons transform into ``primary" hadrons.
This ``hadronization" process is not understood quantitatively and there are
few theoretical predictions for light hadrons, i.e. those that do not contain
a $c$ or $b$ quark.
One approach is to make the ansatz of local parton-hadron duality
(LPHD)~\cite{mlla}, that inclusive distributions of primary hadrons are the
same, up to a normalization factor, as those for partons.
Calculations using MLLA QCD, cut off at a virtual parton mass comparable with
the mass of the hadron in question, have been used in combination with LPHD to
predict properties of the distribution of $\xi=-\ln (2p/E_{CM})$, where $p$ and
$E_{CM}$ are the hadron momentum and the $e^+e^-$ energy, respectively,
in the $e^+e^-$ center of mass frame.
For a given primary hadron species:
the shape should be described well by a Gaussian function within about one unit
of the peak position;
a wider $\xi$ range should be described with the
addition of small distortion terms; and
the peak position should depend inversely on the hadron mass and logarithmically
on $E_{CM}$.
At sufficiently high $x_p = 2p/E_{CM}$, perturbative QCD has also been used to
calculate the $E_{CM}$ dependence of the $x_p$ distributions~\cite{kkp}.
It is desirable to test these predictions experimentally and to encourage deeper
theoretical understanding of the hadronization process.

In the third stage unstable primary hadrons decay into the stable particles that
traverse particle detectors.
This stage is understood inasmuch as proper lifetimes and decay branching
ratios have been measured for many hadron species.
However, these decays complicate fundamental hadronization 
measurements because many of the stable particles are decay
products rather than primary hadrons, and it is typically not possible to
determine whether a given detected hadron is primary.
Previous measurements at $e^+e^-$ colliders (see e.g. \cite{saxon,bohrer})
indicate that decays of vector mesons, strange baryons and decuplet baryons
produce roughly two-thirds of the stable particles;
scalar and tensor mesons, and radially excited baryons have also been
observed \cite{bohrer}, which contribute additional secondaries.
Ideally one would measure every possible hadron species and distinguish
primary hadrons from decay products on a statistical basis.
A body of knowledge could be assembled by
reconstructing heavier and heavier states, and subtracting their known decay
products from the measured differential production rates of lighter hadrons.
The measurements presented here constitute a first step.

Additional complications arise in jets initiated by heavy quarks,
since the leading heavy hadrons carry a large fraction of the beam energy,
restricting that available to other primary hadrons, and
their decays produce a sizable fraction of the stable particles in the jet.
Although the production and decay of some $B$ and $D$ hadrons, containing a $b$
and $c$ quark, respectively, have been studied inclusively~\cite{pdg}
and the effect of heavy quark masses on the charged multiplicity in their jets
has been observed~\cite{chmult,chgxso},
there remain large uncertainties in, e.g., $B^0_s$ and heavy baryon decays,
and heavy quark masses.
The removal of heavy flavor events will therefore allow the direct
study of the hadronization of light quark jets,
and measurements of $c$ and $b$ jets may shed additional light on some of
the above issues.

A particularly interesting aspect of light quark hadronization is the question
of what happens to the quark or antiquark that initiated a given jet.
Recent experimental results~\cite{lpprl,bfp,opallp} have confirmed the notion
that it is ``contained" as a valence constituent of a particular hadron,
and that this ``leading" hadron has on average a higher momentum than the
other hadrons of the same type in the jet.
More precise measurements of such leading particle effects could lead to
methods for identifying jets of a specific light primary flavor,
with a number of applications
in $ep$ and hadron-hadron collisions as well as in $e^+e^-$ annihilations.

There are several phenomenological models of jet production that combine
modelling of all three stages of particle production and it is
important to test their predictions.
To model the parton production stage
the HERWIG~5.8~\cite{herwig}, JETSET~7.4~\cite{jetset74} and
UCLA~4.1~\cite{ucla} event generators use a combination of first order matrix
elements and a parton shower.
To model the hadronization stage, the HERWIG model splits the gluons
produced in the first stage into $q\bar{q}$ pairs, and these quarks and
antiquarks are paired up locally to form colorless clusters that decay into
the primary hadrons.
The JETSET model takes a different approach, representing the color
field between the partons by a semi-classical string;
this string is broken, according to an iterative algorithm,
into several pieces, each of which corresponds to a primary hadron.
In the UCLA model whole events are generated according to weights derived
from the phase space available to their final states and
the relevant Clebsch-Gordan coefficients.
Each of these models contains arbitrary parameters that control various
aspects of hadronization and have been tuned to reproduce data from
$e^+e^-$ annihilations.
The JETSET model includes a large number of parameters that control, on average,
the species of primary hadron produced at each string break,
giving it the potential to model the observed properties of identified hadron
species in great detail.
In the HERWIG model, clusters are decayed into pairs of primary hadrons
according to phase space, and the relative production of different hadrons is
effectively governed by two parameters that control the distribution of cluster
masses.
In the UCLA model, there is only one such free parameter, which
controls the degree of locality of baryon-antibaryon pair formation.

In this paper we present an analysis of inclusive charged particle and
identified $\pi^{\pm}$,
$K^{\pm}$, and p/$\bar{\rm p}$ production in hadronic $Z^0$ decays collected
by the SLC Large Detector (SLD).
The analysis is based upon the approximately 400,000 hadronic events obtained 
in runs of the SLC between 1996 and 1998, and supersedes our previous
measurements~\cite{bfp}.
We measure differential production rates for these hadron
species in a flavor-inclusive sample of hadronic $Z^0$ decays and use
the results to test the predictions of MLLA QCD$+$LPHD and of the three
models just described.
We also measure these differential production rates separately in $Z^0$ decays
into light flavors ($u\bar{u}$, $d\bar{d}$ and $s\bar{s}$), $c\bar{c}$ and
$b\bar{b}$, which provide improved tests of the QCD predictions and new tests of
the models that separate the heavy hadron production and decay
modelling from that of the rest of the hadronization process.
In addition we measure hadron and antihadron differential production rates in
light quark jets, thereby obtaining precise information on leading particles and
additional stringent tests of the models.

In section 2 we describe the SLD, in particular its precision vertexing and
particle identification capabilities.
Section 3 describes the selection of hadronic events, their separation
into samples of different primary flavors using displaced vertices from heavy
hadron decays, and the selection of light quark and antiquark hemispheres
using the large production asymmetry in polar angle induced by the
polarization of the SLC electron beam.
In section 4 we present a measurement of the inclusive stable charged particle
production rate.
We describe the hadron species identification and present
results for flavor-inclusive events in section 5, and
section 6 presents results separately for light-
($Z^0\!\rightarrow\! u\bar{u},d\bar{d},s\bar{s}$),
$c$- ($Z^0\!\rightarrow\! c\bar{c}$) and
$b$-flavor ($Z^0\!\rightarrow\! b\bar{b}$) events.
In sections 7 and 8 we use these results to test the predictions of
MLLA QCD$+$LPHD, and extract total yields of each hadron species per event of
each flavor, respectively.
We present measurements of leading particle production in light-flavor jets
in section 9, and summarize the results in section 10.

\section{The SLD} 

This analysis of data from the SLD~\cite{sld} used charged tracks measured in
the Central Drift Chamber (CDC)~\cite{cdc} and
upgraded CCD Vertex Detector (VXD3)~\cite{vxd3}, and
identified in the Cherenkov Ring Imaging Detector (CRID) \cite{bfp,crid}.
Energy deposits reconstructed in the Liquid Argon Calorimeter (LAC)~\cite{lac}
were used in the initial hadronic event selection and in the calculation of the
event thrust~\cite{thrust} axis.
Momentum measurement was provided by an axial magnetic field of 0.6T.
The CDC and VXD3 gave a momentum resolution of
$\sigma_{p_{\perp}}/p_{\perp}$ = $0.01 \oplus 0.0026p_{\perp}$,
where $p_{\perp}$ is the track momentum transverse to the beam axis in GeV/$c$.

In the plane normal to the beamline the centroid of the micron-sized SLC
interaction region was reconstructed from tracks in sets of approximately
thirty sequential hadronic $Z^0$ decays to a precision of
$\sigma_{IP}^{r\phi}\simeq$3 $\mu$m and used as an estimate of the primary
interaction point (IP).
The IP position along the beam axis was determined event by event using charged
tracks, with an average resolution of $\sigma^z_{IP}\simeq$ 20 $\mu$m.
Including the uncertainty on the IP position, the resolution on the charged
track impact parameter $\delta$ was parametrized in the plane perpendicular
to the beamline as 
$\sigma^{r\phi}_{\delta} =$8$\oplus$29/$(p \sin^{3/2}\theta)$ $\mu$m, and in
any plane containing the beam axis as
$\sigma^z_{\delta} =$9$\oplus$29/$(p \sin^{3/2}\theta)$ $\mu$m,
where $\theta$ is the track polar angle with respect to the beamline. 

The barrel CRID covered the polar angle range $|\cos\theta|<0.68$, and
comprised liquid and gaseous radiator systems with refractive indices of 1.278
and 1.00176, respectively.
In the high momentum limit, an average of 13 (10) Cherenkov photons was
reconstructed per track traversing the liquid (gas) system, with an average
Cherenkov angle resolution of 15 (4.3) mrad.
The combination of these systems gave efficient and pure separation of pions,
kaons and protons over much of the kinematic range, as detailed in section 5.

\section{Event Selection} 

The trigger and initial selection of hadronic events are described
in~\cite{alr}.
The analysis presented here is based on charged tracks measured in the 
CDC and VXD3.
A set of cuts was applied in order to select charged tracks and
events well-contained within the detector acceptance.
Tracks were required to have:
(i) a closest approach to the beam axis within 5 cm, and within
10 cm along the beam axis of the measured IP;
(ii) a polar angle $\theta$
with respect to the beam axis with $|\cos\theta|$ $<$ 0.80;
(iii) a momentum transverse to this axis $p_{\perp}$ $>$
150 MeV/$c$; and
(iv) a momentum $p$ $<$ 50 GeV/c.
Events were required to have:
a minimum of seven such tracks;
a visible energy $E_{vis} > 18$ GeV,
calculated from the accepted tracks, assigned the charged pion mass;
a thrust axis polar angle $\theta_t$ with respect to the beam axis
with $|\cos\theta_t|$ $<$ 0.71; and
the CDC and VXD3 active and a well-measured IP position.
A sample of 284,494 events passed these cuts.
For the identified hadrons, the CRID was also required to be active,
giving a sample of 232,802 events.
The non-hadronic background was estimated to be 0.1\%, dominated by
$Z^0\rightarrow \tau^+ \tau^-$ events.

Samples of events enriched in light, $c$, and $b$ primary flavors were
selected~\cite{hjkang} using tracks with well measured impact parameters.
For each event we defined $n_{sig}$ as the number of tracks with an impact
parameter greater than three times its estimated error,
$\delta > 3 \sigma_{\delta}$.
We also ran a topological vertex finding algorithm~\cite{davej} on the set of
tracks in each hemisphere and considered the $p_t$-corrected mass~\cite{davej}
$M_{pt}$ of any secondary vertex (significantly separated from the IP) found.
Any event containing a vertex with $M_{pt}>2$ GeV/c$^2$ was assigned to the
$b$-tagged sample.
If not $b$-tagged, an event was $c$-tagged if either:
it contained a vertex with 0.5$<\!\!M_{pt}\!\!<$2 GeV/c$^2$,
a total momentum of tracks assigned to the vertex $P_{vtx}\!\!>$2 GeV/c
and $cP_{vtx}-14c^2M_{pt}\!\!>\!\!-10$ GeV;
or it contained no secondary vertex and had $2 \leq n_{sig} \leq 3$.
Events with no secondary vertex and $n_{sig}=0$ 
were assigned to the light-tagged sample.
The remaining events were kept as an untagged sample.
The 15\% of the data taken in 1996 were excluded due to uncertainties in the
simulation, and
the light-, $c$- and $b$-tagged samples comprised roughly
83,000, 28,000 and 33,000 events, respectively.
Selection efficiencies were calculated from a detailed detector simulation
based on a version of JETSET tuned to the world's data on inclusive particle
production and $D$ and $B$ hadron production and decay~\cite{sldsim};
efficiencies and sample purities for the above selection are listed in
table~\ref{tlveff}.

\begin{table}
\begin{center}
 \begin{tabular}{|r||c|c|c||c|c|c|} \hline
       & \multicolumn{3}{c||}{ } & \multicolumn{3}{c|}{ } \\ [-.3cm]
       & \multicolumn{3}{c||}{Efficiency for $Z^0 \rightarrow$}
       & \multicolumn{3}{c|}{Purity of $Z^0 \rightarrow$} \\
       & $u\bar{u},d\bar{d},s\bar{s}$ & $c\bar{c}$ & $b\bar{b}$ 
       & $u\bar{u},d\bar{d},s\bar{s}$ & $c\bar{c}$ & $b\bar{b}$  \\ [.1cm] \hline 
 &&&&&&\\[-.3cm] 
$uds$-tag  & 0.703 & 0.190 & 0.010 & 0.925 & 0.070 & 0.005 \\
$c$-tag    & 0.061 & 0.551 & 0.105 & 0.204 & 0.638 & 0.158 \\
$b$-tag    & 0.001 & 0.024 & 0.815 & 0.005 & 0.023 & 0.973 \\
 no-tag    & 0.235 & 0.235 & 0.070 & 0.721 & 0.201 & 0.078 \\[.1cm] \hline 
 \end{tabular}
\caption{\baselineskip=12pt \label{tlveff}
Efficiencies for simulated events in the three flavor categories to be
tagged as light ($uds$), $c$ or $b$ flavor, or none of these.
The three rightmost columns indicate the composition of each simulated tagged
sample assuming Standard Model relative flavor production.}
\end{center}
\end{table}

Samples of hemispheres enriched in light-quark and light-antiquark jets
were selected by exploiting the large electroweak forward-backward production
asymmetry with respect to the beam direction induced by the high polarization of
the SLC electron beam.
Here a looser light-flavor event tag of $n_{sig}=0$ was applied, giving a
simulated $uds$ efficiency of 84\% and purity of 89\%, with the background
dominated by $c$-flavor events.
The event thrust axis was used to approximate the initial $q\bar{q}$ axis and
was signed such that its $z$-component was along the electron beam direction,
$\hat{t}_z>0$.
Events in the central region of the detector, where the production asymmetry is
small, were removed by the requirement $|\hat{t}_z|>0.15$, leaving 109,000
events.
The quark-tagged hemisphere in events with left- (right-)handed electron beam
polarization was defined to comprise the set of tracks with positive (negative)
momentum projection along the signed thrust axis.
The remaining tracks in each event
were defined to be in the antiquark-tagged hemisphere.
For the selected event sample, the average magnitude of the $e^-$ beam
polarization was 0.73.
Using this value and assuming Standard Model couplings,
a tree-level calculation gives a quark (antiquark) purity of 0.73 in the
quark-(antiquark-)tagged sample.

\section{Stable Charged Particle Rates} 

We first measured the differential production rate of all stable
charged particles defined conventionally as the sum of electrons, muons,
pions, kaons and protons that are either primary hadrons or products of a chain
of decays of particles with proper lifetime less than 3$\times10^{-10}$s.
Tracks satisfying the requirements in sec. 3 were counted in momentum bins and
corrected, using our detector simulation, for the track finding and selection
efficiency and resolution, non-hadronic event background, and spurious tracks
from beam-related backgrounds and interactions in the detector material.
The resulting integrated rate was constrained to be 20.95$\pm$0.21
charged particles per event, an average of
measurements~\cite{chgxso,chgxsa,chgxsd} in $Z^0$ decays.

The momentum dependence of the selection was constrained by comparing
the properties of reconstructed charged tracks in measured and simulated
$\tau$ lepton decays,
for which the momentum and particle type distributions are well known.
We selected $e^+e^-\rightarrow \tau^+\tau^-$ events according to
the criteria in \cite{alept}, with the additional
requirement of at least one track in the event with $p>$7 GeV/c, to reduce
beam-related and two-photon event background.
Comparisons were made for 1-, 3-, and 5-prong decays separately, as well as for
hemispheres in which 2 or 4 tracks were found, giving constraints on both
isolated tracks and those in close proximity to others.
We also checked the momentum distributions for tagged electrons, muons, pions,
and kaons, and the overall multiplicity distribution in selected $\tau$-pair
events.
In all cases the simulation was consistent with the data.
Linear fits were made to the ratios of data:simulated tracks,
and a momentum-dependent uncertainty of 0.10$\times |p-2.7$ GeV/c$|$\% was
assigned, reflecting the error on a typical fitted slope.
Here $p$ is the particle momentum in GeV/c and 2.7 GeV/c is the average
momentum of all charged particles, so that this uncertainty is independent of
the normalization uncertainty noted above.
Variations of the background levels, detector material and momentum resolution
were also considered and found to have much smaller effects, except at the
lowest and highest momenta.

\begin{figure}
 \vspace{-1.2cm}
  \epsfxsize=6.5in
  \begin{center}\mbox{\epsffile{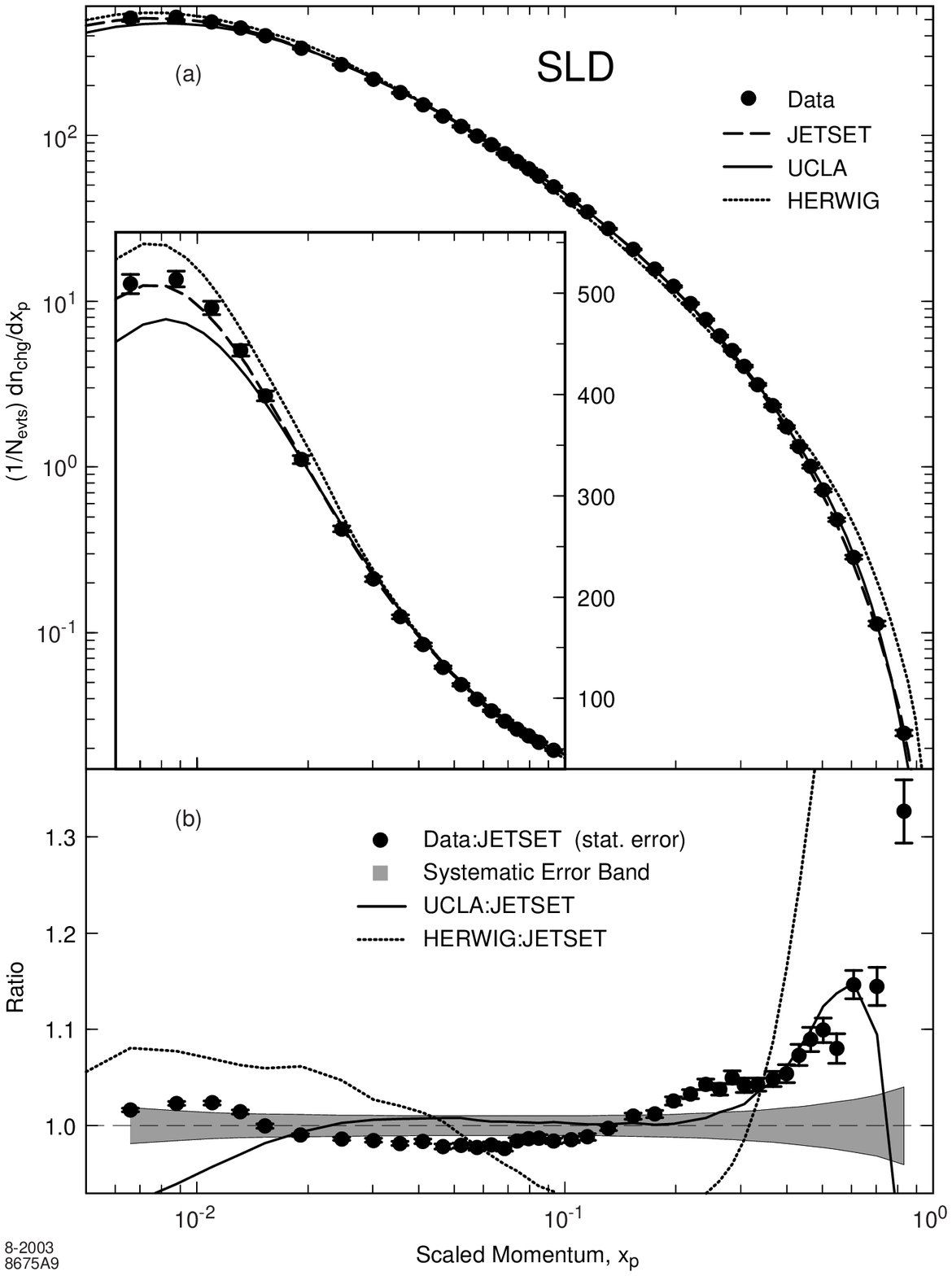}}\end{center}
 \vspace{-0.8cm}
 \caption{ \baselineskip=12pt  \label{chgxsall}
a) Charged track $x_p$ distribution in hadronic events (dots)
with logarithmic and (inset) linear vertical scales,
compared with the predictions of three models (lines).
b) The same data divided by the JETSET prediction.
The error bars in (b) are statistical and the shaded band
represents the systematic uncertainty; all errors 
except an overall 1\% normalization uncertainty are included in (a).
    }
\end{figure}

The inclusive charged particle differential production rate is listed in
table~\ref{chgxs}, in terms of the variables momentum $p$,
scaled momentum $x_p=2p/E_{CM}$ and $\xi=-\ln x_p$.
The $x_p$ distribution is shown in fig.~\ref{chgxsall},
and compared with the predictions of the JETSET, UCLA and HERWIG event
generators described in section 1, using the default parameter values for
each model.
The JETSET model is the most consistent with the data, but predicts a slightly
softer spectrum;
the UCLA model describes the data over most of the $x_p$ region, but falls low
for $x_p\! >$0.7 and $x_p\! <$0.015 units;
the HERWIG model predicts a spectrum quite different from that seen in the data.
Our results for the shape of the spectrum are consistent with those published
previously~\cite{chgxso,chgxsa,chgxsd}.

\begin{table}
\begin{center}
\begin{tabular}{|r@{--}r|c|r@{$\pm$}l@{$\pm$}l@{$\pm$}l|
                          |r@{$\pm$}l||r@{$\pm$}l|} \hline
\multicolumn{11}{|c|}{ } \\ [-.45cm]
\multicolumn{11}{|c|}{Inclusive Stable Charged Particle Production Rate}\\[.03cm]\hline
\multicolumn{2}{|c|}{ } && \multicolumn{4}{|c||}{ } &
\multicolumn{2}{|c||}{ }  & \multicolumn{2}{|c|}{ }\\[-.45cm]
\multicolumn{2}{|c|} {$p$ (GeV/c)} & $\left< p \right>$ &
$\frac{1}{N_{evt}} \frac{dn_{chg}}{dp}$   & stat. & effic. & other &
\multicolumn{2}{|c||}{$\frac{1}{N_{evt}} \frac{dn_{chg}}{dx_p}$} &
\multicolumn{2}{|c|} {$\frac{1}{N_{evt}} \frac{dn_{chg}}{d\xi}$} \\ [.13cm]\hline
0.25& 0.35&0.300&11.166&0.022&0.115&0.173& 509.2&9.5  & 3.234&0.062\\[-0.03cm]
0.35& 0.45&0.400&11.271&0.022&0.116&0.128& 513.9&7.8  & 4.399&0.068\\[-0.03cm]
0.45& 0.55&0.499&10.646&0.022&0.109&0.096& 485.5&6.6  & 5.221&0.072\\[-0.03cm]
0.55& 0.65&0.599& 9.724&0.021&0.099&0.073& 443.4&5.6  & 5.665&0.073\\[-0.03cm]
0.65& 0.75&0.699& 8.744&0.020&0.089&0.057& 398.7&4.9  & 6.110&0.075\\[-0.03cm]
0.75& 1.00&0.870& 7.364&0.011&0.075&0.040& 335.8&3.9  & 6.324&0.074\\[-0.03cm]
1.00& 1.25&1.121& 5.860&0.010&0.059&0.024& 267.2&2.9  & 6.486&0.072\\[-0.03cm]
1.25& 1.50&1.371& 4.775&0.009&0.048&0.015& 217.8&2.3  & 6.465&0.070\\[-0.03cm]
1.50& 1.75&1.622& 3.962&0.008&0.040&0.011& 180.7&1.9  & 6.338&0.068\\[-0.03cm]
1.75& 2.00&1.872& 3.360&0.008&0.034&0.009& 153.2&1.6  & 6.202&0.066\\[-0.03cm]
2.00& 2.25&2.122& 2.865&0.007&0.029&0.007& 130.6&1.4  & 5.990&0.063\\[-0.03cm]
2.25& 2.50&2.375& 2.488&0.007&0.025&0.005& 113.4&1.2  & 5.807&0.062\\[-0.03cm]
2.50& 2.75&2.625& 2.173&0.006&0.022&0.004& 99.09&1.03 & 5.605&0.059\\[-0.03cm]
2.75& 3.00&2.875& 1.920&0.006&0.019&0.004& 87.57&0.91 & 5.419&0.058\\[-0.03cm]
3.00& 3.25&3.125& 1.701&0.005&0.017&0.003& 77.56&0.81 & 5.214&0.056\\[-0.03cm]
3.25& 3.50&3.375& 1.530&0.005&0.015&0.003& 69.78&0.73 & 5.061&0.054\\[-0.03cm]
3.50& 3.75&3.625& 1.378&0.005&0.014&0.002& 62.84&0.66 & 4.892&0.053\\[-0.03cm]
3.75& 4.00&3.875& 1.244&0.005&0.013&0.002& 56.72&0.60 & 4.713&0.051\\[-0.03cm]
4.00& 4.50&4.244& 1.072&0.003&0.011&0.001& 48.90&0.51 & 4.450&0.047\\[-0.03cm]
4.50& 5.00&4.744& 0.894&0.003&0.009&0.001& 40.78&0.43 & 4.141&0.045\\[-0.03cm]
5.00& 5.50&5.244& 0.754&0.003&0.008&0.001& 34.39&0.37 & 3.856&0.042\\[-0.03cm]
5.50& 6.50&5.975& 0.600&0.002&0.006&0.001& 27.35&0.29 & 3.492&0.038\\[-0.03cm]
6.50& 7.50&6.977&0.4502&0.0014&0.0049&0.0005& 20.53&0.23 & 3.051&0.035\\[-0.03cm]
7.50& 8.50&7.980&0.3429&0.0012&0.0039&0.0003& 15.64&0.18 & 2.647&0.032\\[-0.03cm]
8.50& 9.50&8.982&0.2687&0.0011&0.0032&0.0003& 12.25&0.15 & 2.328&0.029\\[-0.03cm]
9.50&10.50&9.980&0.2120&0.0010&0.0026&0.0002&  9.67&0.12 & 2.039&0.027\\[-0.03cm]
10.5& 11.5&10.99&0.1700&0.0009&0.0022&0.0002&  7.75&0.11 & 1.792&0.025\\[-0.03cm]
11.5& 12.5&11.99&0.1351&0.0008&0.0018&0.0001& 6.161&0.088& 1.551&0.023\\[-0.03cm]
12.5& 13.5&12.99&0.1103&0.0007&0.0016&0.0001& 5.029&0.076& 1.368&0.022\\[-0.03cm]
13.5& 14.5&13.99&0.0889&0.0006&0.0013&0.0001& 4.053&0.065& 1.184&0.020\\[-0.03cm]
14.5& 16.0&15.22&0.0688&0.0005&0.0011&0.0001& 3.139&0.052& 0.997&0.017\\[-0.03cm]
16.0& 17.5&16.72&0.0513&0.0004&0.0009&0.0001& 2.338&0.042& 0.812&0.016\\[-0.03cm]
17.5& 19.0&18.23&0.0383&0.0003&0.0007&0.0001& 1.748&0.034& 0.661&0.014\\[-0.03cm]
19.0& 20.5&19.72&0.0291&0.0003&0.0006&0.0001& 1.326&0.028& 0.544&0.012\\[-0.03cm]
20.5& 22.0&21.21&0.0221&0.0003&0.0005&0.0001& 1.008&0.023& 0.444&0.011\\[-0.03cm]
22.0& 24.0&22.96&0.0159&0.0002&0.0004&0.0001& 0.724&0.018&0.3468&.0090\\[-0.03cm]
24.0& 26.0&24.94&0.0105&0.0001&0.0003&0.0001& 0.480&0.013&0.2513&.0073\\[-0.03cm]
26.0& 30.0&27.74&0.0063&0.0001&0.0002&0.0001& 0.285&0.009&0.1700&.0056\\[-0.03cm]
30.0& 35.0&32.14&0.0025&0.0000&0.0001&0.0001& 0.114&0.005&0.0800&.0035\\[-0.03cm]
35.0& 45.0&38.23&0.0005&0.0000&0.0000&0.0001& 0.024&0.001&0.0207&.0013
 \\[.03cm] \hline 
 \end{tabular}
\caption{\baselineskip=12pt \label{chgxs}
Production rate for all stable charged particles in terms of momentum $p$,
scaled momentum $x_p$ and $\xi=-\ln x_p$.
For momentum we give statistical errors and the systematics arising from
track-finding efficiency (including an overall 1\% normalization term) and the
sum of other sources, which is dominated by backgrounds (momentum resolution)
at low (high) momentum;
in the other columns these have been summed in quadrature.}
\end{center}
\end{table}

\section{Charged Pion, Kaon and Proton Production} 

Additional track selection cuts \cite{hjkang} were applied to remove tracks
that interacted or scattered through large angles before exiting the CRID and 
to ensure that the CRID performance was well-modelled by the simulation.
Good information from the liquid (gas) system was required for tracks in the
``liquid" (``gas") region, with momenta below 2.25 GeV/c (above 10.5 GeV/c);
good information from both systems was required for momenta in the ``combined"
region, 2.25--10.5 GeV/c.
Tracks were required to have at least 40 CDC hits, at least one of which
was at a radius of at least 92 cm,
to extrapolate through an active region of the appropriate radiator(s),
and
to have at least 40 (70)\% of their expected liquid (gas) ring contained
within a sensitive region of the CRID TPCs.
The latter requirement included dead and inefficient regions within the TPCs
and also rejected tracks with $p>2.25$ GeV/c for which there was a saturated
CRID hit within a 2 cm radius ($\sim$the maximum ring radius) of a point 1 cm
behind the expected gas ring center in the photoelectron drift direction.
Saturated hits arise from the passage of tracks through the TPCs and from
various backgrounds, and mask single photoelectron hits in their vicinity.
Liquid region tracks that extrapolated through an active TPC were required to
have a saturated hit within 1 cm of the extrapolated track, indicating that the
track traversed the TPC and hence the liquid radiator;
gas region tracks were required to have either such a saturated hit or
the presence of at least four hits consistent with a liquid ring.
These cuts accepted 71\%, 68\% and 74\% of the tracks within the CRID
geometrical acceptance
in the liquid, combined and gas regions, respectively.

Tracks were identified using a global likelihood technique \cite{bfp,davea}. 
For each track and each of the five hypotheses $i=e,\mu,\pi,K$, p, a likelihood
$L_i$ was calculated based upon:
the number of detected photoelectrons and their measured Cherenkov angles;
the expected number of photons;
the expected Cherenkov angle; and 
terms accounting for random backgrounds and hits consistent with Cherenkov
radiation from other tracks in the event.
The best hypothesis for each track was used to determine its contribution to the
background for other tracks, and the calculation was iterated until there was
no change in any best hypothesis.
Particle separation was based upon differences between logarithms of the three
likelihoods, ${\cal L}_i = \ln L_i$, $i=\pi,K$, p.
A track in the liquid (combined, gas) region was identified as species $j$
if ${\cal L}_j$ exceeded both of the other log-likelihoods by at least 3 (2)
units.
The electron and muon likelihoods are generally quite similar to the pion
likelihood, and the leptons were included in the pion category at this stage.

\begin{figure}
 \vspace{-0.2cm}
  \epsfxsize=6.7in
  \begin{center}\mbox{\epsffile{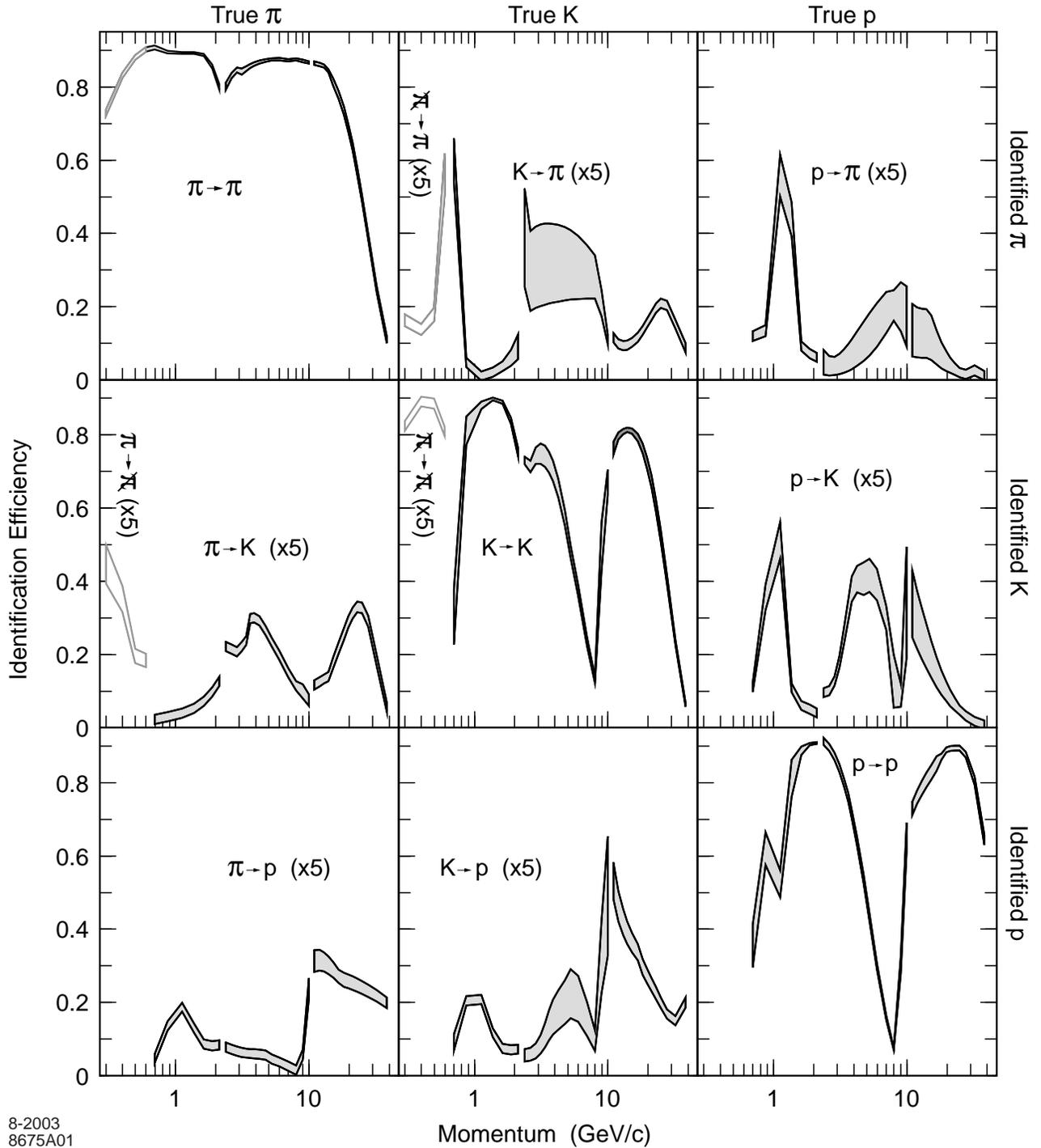}}\end{center}
 \vspace{-0.8cm}
 \caption{ \baselineskip=12pt  \label{effpar}
Calibrated identification efficiencies for selected tracks.
The half-widths of the grey bands represent the systematic
uncertainties, which are strongly correlated between momenta.
The off-diagonal efficiencies have been scaled by factors of five for clarity.
The white bands in the four upper left plots represent the 2$\times$2 matrix
used below kaon threshold in the liquid system.
    }
\end{figure}

We quantified the identification performance in terms of a momentum-dependent
identification efficiency matrix {\bf E}, each element $E_{ij}$ of which
represents the probability that the selected track from a true $i$-hadron is
identified as a $j$-hadron, with $i,j=\pi$,$K$,p.
The elements of this matrix were determined where possible from the data
\cite{hjkang,tomp}.
For example, tracks from selected $K_s^0$ and $\tau$ decays were used as
``pion" test samples, having estimated kaon plus proton contents of 0.3\%
and 1.7\% respectively.
Probabilities for such tracks to be identified as pions, kaons and protons were
compared with the predictions of our detailed MC simulation, which
was found to describe the momentum dependence of the efficiencies well
and to reproduce their amplitudes to within a few percent.
Functional forms,
chosen to describe the momentum dependence of both data and simulated test
samples as well as that of simulated true pions in hadronic events,
were fitted to the data,
except for momenta below 0.8 GeV/c where there was significant structure on the
scale of the bin size and bin-by-bin corrections were used.
The simulation was used to correct the fitted parameters for non-pion content in
the test samples and differences in tracking performance between tracks
in these samples and those passing our selection cuts in hadronic events.
The resulting identification efficiency functions, $E_{\pi \pi}$, $E_{\pi K}$
and $E_{\pi {\rm p}}$, are shown in the leftmost column of fig. \ref{effpar}.
A similar procedure using only $\pi$ and p likelihoods was used to measure the
$\pi$-p separation in the liquid (gas) system for $p>2$ (17) GeV/c.
This information was combined with that from test samples of protons and kaons
from decays of $\Lambda^0$ hyperons and $\phi$ mesons, respectively, to derive
the remaining efficiencies in fig. \ref{effpar}.

The bands in fig. \ref{effpar} encompass the upper and lower 
systematic error bounds on the efficiencies.
There are discontinuities between the liquid, combined and gas regions, and
there are strong point-to-point correlations within each region.
For the diagonal elements, these errors correspond to those on the fitted
parameters, and fall into four categories.
The overall amplitude of the efficiency is driven by the average and rms of
the number of photons detected per track;  the corresponding uncertainty is
common to all momenta in a given region and is also correlated between the
liquid and combined regions for protons and the
combined and gas regions for pions.
The Cherenkov angle resolution affects the positions of the falling edges;
its uncertainty is therefore correlated across the range of a given edge, as
well as between pions and kaons in the regions 1.5--2.25 GeV/c and
12--45 GeV/c, and between kaons and protons in the region 3--8 GeV/c.
Performance near a Cherenkov threshold depends on the relevant index of
refraction and its stability;  since the efficiencies change rapidly on the
scale of our bin sizes, bin-by-bin calibrations were done for
$E_{KK}$ ($E_{\rm pp}$) in the range $p<$1 GeV/c ($p<1.5$ GeV/c),
and $E_{KK}$ and $E_{\rm pp}$ in the range 7.5$<p<$10.5 GeV/c, which are
completely independent.
In the region 10.5--18 GeV/c, protons are below threshold in the gas, whereas
pions and kaons are well above threshold;  here $E_{\rm pp}$ depends largely on
the background level, and its uncertainty is correlated across this region.

For the off-diagonal elements, representing misidentification rates, the errors
on fitted parameters were also used, but subject to a minimum value of $0.0025$
to account for the limited statistics of the test sample
constraints on the momentum dependences.
These uncertainties correspond to a combination of the effects listed above for
the diagonal elements, and each is typically dominated by one effect in a given
momentum region.
These errors should also therefore be considered strongly positively correlated
across each of the liquid, combined and gas regions.

The identification efficiencies in fig. \ref{effpar} peak near or above
0.9 and are greater than 0.8 over wide ranges.
The pion (kaon and proton) coverage is continuous from 0.25 GeV/c (0.65 GeV/c)
up to the beam energy, although the efficiencies fall below 0.2 for pions and
kaons above about 30 GeV/c, and kaons and protons in the 6--9 GeV/c range.
Misidentification rates are typically at the few percent level, with peak
values of up to 0.1. 

\begin{figure}
 \vspace{-0.2cm}
  \epsfxsize=6.5in
  \begin{center}\mbox{\epsffile{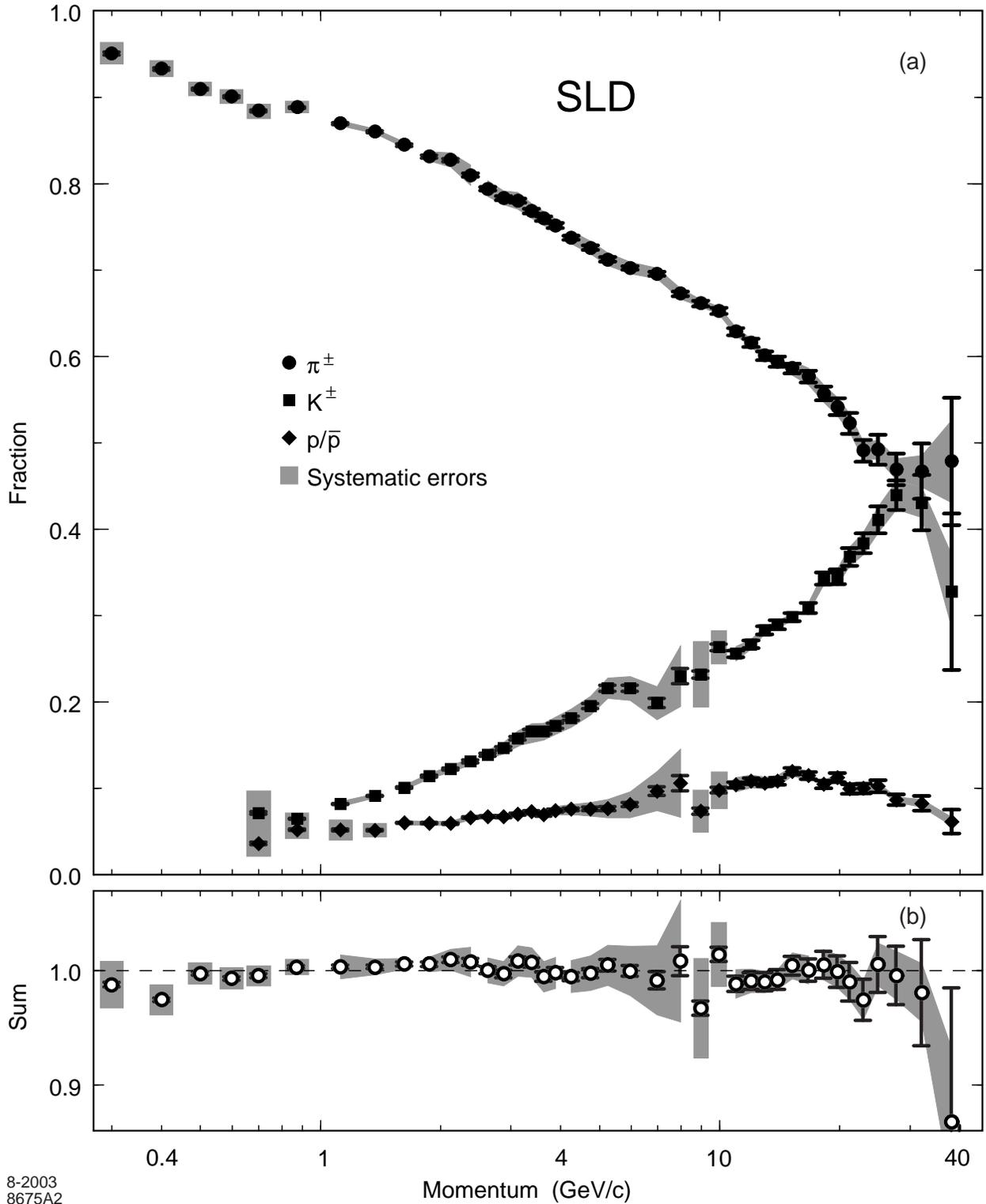}}\end{center}
 \vspace{-0.8cm}
 \caption{ \baselineskip=12pt  \label{fraxg}
a) Measured charged hadron production fractions in hadronic $Z^0$ decays. 
The circles represent the $\pi^\pm$ fraction, the squares the $K^\pm$ fraction,
the diamonds the p/$\bar{\rm p}$ fraction.
The error bars are statistical only;
the shaded areas indicate the systematic errors, and are connected across
momentum regions where there is a strong positive correlation.
b) The sum of the three fractions.
    }
\end{figure}

In each momentum bin we measured the fractions of the selected tracks
that were identified as pions, kaons and protons.
The observed fractions were related to the true production fractions by an
efficiency matrix, composed of the values shown in fig.~\ref{effpar}.
This matrix was inverted and used to unfold our observed identified hadron
fractions.
This analysis procedure does not require that the sum 
of the charged hadron fractions be unity; instead the sum was used as a 
consistency check, which was found to be satisfied at all momenta (see fig.
\ref{fraxg}). 
For momenta below 0.65 GeV/c, we could not distinguish kaons from protons, but
pions could be identified down to 0.25 GeV/c.
An analogous 2$\times$2 analysis of pions and non-pions was used in this region,
and we present only the pion fraction.

The background from electrons and muons was estimated from the simulation to
be about 5\% of the tracks in the inclusive flavor sample,
predominantly from $c$- and $b$-flavor events.
The fractions were corrected for the lepton backgrounds
using the simulation,
as well as for the effects of beam-related
backgrounds, particles interacting in the detector material, and particles
decaying outside the tracking volume.
The conventional definition of a final-state charged hadron was used,
namely a charged pion, kaon or proton that is either
from the primary interaction or a product of a chain of decays of hadrons each
with a proper lifetime less than 3$\times10^{-10}$s.

The measured charged hadron fractions in inclusive hadronic $Z^0$ decays
are shown in fig.~\ref{fraxg} and listed in tables \ref{pifraxa}--\ref{pfraxa}.
The systematic errors were determined by propagating the errors on the
calibrated efficiency matrix and are strongly correlated as described above.
They are indicated by the shaded regions in fig. \ref{fraxg}.
The errors on the points below $\sim$15 GeV/c are dominated by the systematic
uncertainties;
those above $\sim$30 GeV/c are dominated by statistical uncertainties.
The sum of fractions is consistent with unity everywhere.

\begin{table} 
\begin{center}
\begin{tabular}{|r@{--}r|c|r@{$\pm$}r@{$\pm$}r|r@{$\pm$}l@{$\pm$}l|} \hline
\multicolumn{2}{|c|}{$x_p$ Range} & $<\! x_p \!>$ &
\multicolumn{3}{c|}{$f_{\pi}$} &
\multicolumn{3}{c|}{(1/N$_{evts}$) d$n_{\pi}$/d$x_p$}
\\ \hline
0.005&0.008 &0.0066 & 0.951 &0.002 &0.010 & 471.8 &   1.3 &   9.2 \\[-0.03cm]
0.008&0.010 &0.0088 & 0.933 &0.001 &0.007 & 470.4 &   1.1 &   6.6 \\[-0.03cm]
0.010&0.012 &0.0109 & 0.910 &0.001 &0.006 & 434.6 &   1.1 &   5.0 \\[-0.03cm]
0.012&0.014 &0.0131 & 0.901 &0.001 &0.006 & 388.8 &   1.0 &   4.0 \\[-0.03cm]
0.014&0.016 &0.0153 & 0.885 &0.001 &0.006 & 352.7 &   0.9 &   3.3 \\[-0.03cm]
0.016&0.022 &0.0191 & 0.888 &0.001 &0.004 & 294.8 &   0.5 &   2.2 \\[-0.03cm]
0.022&0.027 &0.0246 & 0.870 &0.001 &0.003 & 229.6 &   0.5 &   1.3 \\[-0.03cm]
0.027&0.033 &0.0301 & 0.860 &0.001 &0.003 & 185.0 &   0.4 &   0.9 \\[-0.03cm]
0.033&0.038 &0.0356 & 0.845 &0.001 &0.003 & 150.6 &   0.4 &   0.7 \\[-0.03cm]
0.038&0.044 &0.0411 & 0.831 &0.002 &0.005 & 125.6 &   0.4 &   0.9 \\[-0.03cm]
0.044&0.049 &0.0465 & 0.828 &0.002 &0.008 & 106.5 &   0.4 &   1.1 \\[-0.03cm]
0.049&0.055 &0.0521 & 0.810 &0.002 &0.012 & 90.40 &  0.35 &  1.33 \\[-0.03cm]
0.055&0.060 &0.0576 & 0.794 &0.002 &0.009 & 77.38 &  0.31 &  0.91 \\[-0.03cm]
0.060&0.066 &0.0630 & 0.784 &0.002 &0.008 & 67.39 &  0.29 &  0.70 \\[-0.03cm]
0.066&0.071 &0.0685 & 0.780 &0.002 &0.010 & 59.40 &  0.27 &  0.75 \\[-0.03cm]
0.071&0.077 &0.0740 & 0.768 &0.003 &0.009 & 52.57 &  0.25 &  0.60 \\[-0.03cm]
0.077&0.082 &0.0795 & 0.760 &0.003 &0.008 & 46.76 &  0.24 &  0.50 \\[-0.03cm]
0.082&0.088 &0.0850 & 0.752 &0.003 &0.008 & 41.70 &  0.23 &  0.43 \\[-0.03cm]
0.088&0.099 &0.0931 & 0.738 &0.002 &0.007 & 35.26 &  0.15 &  0.36 \\[-0.03cm]
0.099&0.110 &0.1040 & 0.726 &0.002 &0.007 & 28.89 &  0.13 &  0.29 \\[-0.03cm]
0.110&0.121 &0.1150 & 0.712 &0.003 &0.007 & 23.88 &  0.12 &  0.25 \\[-0.03cm]
0.121&0.143 &0.1310 & 0.702 &0.002 &0.007 & 18.69 &  0.08 &  0.19 \\[-0.03cm]
0.143&0.164 &0.1530 & 0.696 &0.002 &0.006 & 13.85 &  0.07 &  0.14 \\[-0.03cm]
0.164&0.186 &0.1750 & 0.673 &0.003 &0.006 & 10.16 &  0.06 &  0.11 \\[-0.03cm]
0.186&0.208 &0.1970 & 0.662 &0.003 &0.004 & 7.812 & 0.050 & 0.069 \\[-0.03cm]
0.208&0.230 &0.2189 & 0.653 &0.004 &0.004 & 6.076 & 0.044 & 0.061 \\[-0.03cm]
0.230&0.252 &0.2410 & 0.629 &0.004 &0.005 & 4.674 & 0.039 & 0.053 \\[-0.03cm]
0.252&0.274 &0.2629 & 0.616 &0.005 &0.005 & 3.632 & 0.035 & 0.044 \\[-0.03cm]
0.274&0.296 &0.2849 & 0.601 &0.005 &0.004 & 2.886 & 0.031 & 0.037 \\[-0.03cm]
0.296&0.318 &0.3068 & 0.594 &0.006 &0.004 & 2.292 & 0.028 & 0.031 \\[-0.03cm]
0.318&0.351 &0.3338 & 0.586 &0.006 &0.009 & 1.749 & 0.021 & 0.034 \\[-0.03cm]
0.351&0.384 &0.3666 & 0.577 &0.007 &0.010 & 1.275 & 0.018 & 0.028 \\[-0.03cm]
0.384&0.417 &0.3997 & 0.557 &0.008 &0.010 & 0.921 & 0.016 & 0.022 \\[-0.03cm]
0.417&0.450 &0.4325 & 0.542 &0.010 &0.010 & 0.680 & 0.014 & 0.018 \\[-0.03cm]
0.450&0.482 &0.4651 & 0.523 &0.012 &0.011 & 0.499 & 0.013 & 0.014 \\[-0.03cm]
0.482&0.526 &0.5035 & 0.491 &0.013 &0.010 & 0.338 & 0.010 & 0.010 \\[-0.03cm]
0.526&0.570 &0.5470 & 0.492 &0.018 &0.011 & 0.226 & 0.009 & 0.007 \\[-0.03cm]
0.570&0.658 &0.6083 & 0.469 &0.018 &0.012 & 0.130 & 0.005 & 0.005 \\[-0.03cm]
0.658&0.768 &0.7047 & 0.467 &0.032 &0.018 &0.0526 &0.0037 &0.0029 \\[-0.03cm]
0.768&1.000 &0.8383 & 0.479 &0.074 &0.048 &0.0113 &0.0018 &0.0013
 \\ \hline \hline 
0.005&1.000 & \multicolumn{4}{c|}{  }& 15.74 & 0.01 & 0.17
 \\ \hline 
 \end{tabular}
\vspace{-0.2cm}
\caption{\baselineskip=12pt  \label{pifraxa}
Charged pion fraction $f_{\pi}$ and differential production rate
(1/N$_{evts}$)d$n_{\pi}$/d$x_p$ per hadronic $Z^0$ decay.
The first error is statistical, the second systematic.
$<\!\! x_p\!\!>$ is the average $x_p$-value of charged tracks in each bin.
The last row gives the integral over the $x_p$ range of the measurement.
A 1.0\% normalization uncertainty is included in the systematic error on the
integral, but not in those on the differential rate.}
\end{center}
\end{table}

\begin{table}
\begin{center}
 \begin{tabular}{|c|c|r@{$\pm$}r@{$\pm$}r|r@{$\pm$}l@{$\pm$}l|} \hline
 $x_p$ Range & $<\! x_p\! >$ & \multicolumn{3}{c|}{$f_K$}  &
      \multicolumn{3}{c|}{(1/N$_{evts}$) d$n_{K}$/d$x_p$}   \\ \hline 
  & & \multicolumn{3}{c|}{ } & \multicolumn{3}{c|}{  }   \\[-.3cm]
0.014--0.016 &0.0153 & 0.072 &0.002 &0.023 & 28.59 &  0.64 &  9.26 \\
0.016--0.022 &0.0191 & 0.065 &0.001 &0.005 & 21.57 &  0.20 &  1.57 \\
0.022--0.027 &0.0246 & 0.082 &0.001 &0.003 & 21.62 &  0.19 &  0.80 \\
0.027--0.033 &0.0301 & 0.091 &0.001 &0.002 & 19.65 &  0.18 &  0.53 \\
0.033--0.038 &0.0356 & 0.101 &0.001 &0.002 & 18.02 &  0.16 &  0.44 \\
0.038--0.044 &0.0411 & 0.114 &0.001 &0.003 & 17.27 &  0.17 &  0.43 \\
0.044--0.049 &0.0465 & 0.123 &0.001 &0.004 & 15.78 &  0.17 &  0.47 \\
0.049--0.055 &0.0521 & 0.131 &0.002 &0.004 &14.664 & 0.194 & 0.442 \\
0.055--0.060 &0.0576 & 0.139 &0.002 &0.005 &13.535 & 0.189 & 0.503 \\
0.060--0.066 &0.0630 & 0.147 &0.002 &0.006 &12.599 & 0.176 & 0.558 \\
0.066--0.071 &0.0685 & 0.158 &0.002 &0.008 &12.036 & 0.165 & 0.635 \\
0.071--0.077 &0.0740 & 0.166 &0.002 &0.009 &11.349 & 0.162 & 0.622 \\
0.077--0.082 &0.0795 & 0.166 &0.003 &0.010 &10.207 & 0.164 & 0.603 \\
0.082--0.088 &0.0850 & 0.172 &0.003 &0.010 & 9.571 & 0.160 & 0.566 \\
0.088--0.099 &0.0931 & 0.181 &0.002 &0.011 & 8.671 & 0.113 & 0.505 \\
0.099--0.110 &0.1040 & 0.196 &0.003 &0.011 & 7.784 & 0.114 & 0.440 \\
0.110--0.121 &0.1150 & 0.216 &0.004 &0.012 & 7.237 & 0.120 & 0.395 \\
0.121--0.143 &0.1310 & 0.216 &0.003 &0.014 & 5.746 & 0.089 & 0.369 \\
0.143--0.164 &0.1530 & 0.199 &0.005 &0.019 & 3.959 & 0.102 & 0.381 \\
0.164--0.186 &0.1750 & 0.230 &0.009 &0.035 & 3.473 & 0.134 & 0.532 \\
0.186--0.208 &0.1970 & 0.232 &0.004 &0.035 & 2.739 & 0.047 & 0.419 \\
0.208--0.230 &0.2189 & 0.264 &0.004 &0.017 & 2.452 & 0.037 & 0.163 \\
0.230--0.252 &0.2410 & 0.256 &0.004 &0.008 & 1.903 & 0.030 & 0.063 \\
0.252--0.274 &0.2629 & 0.267 &0.004 &0.006 & 1.574 & 0.027 & 0.036 \\
0.274--0.296 &0.2849 & 0.283 &0.005 &0.004 & 1.360 & 0.024 & 0.026 \\
0.296--0.318 &0.3068 & 0.290 &0.005 &0.004 & 1.118 & 0.022 & 0.020 \\
0.318--0.351 &0.3338 & 0.298 &0.005 &0.004 & 0.890 & 0.016 & 0.017 \\
0.351--0.384 &0.3666 & 0.309 &0.006 &0.006 & 0.683 & 0.014 & 0.016 \\
0.384--0.417 &0.3997 & 0.343 &0.007 &0.008 & 0.567 & 0.013 & 0.015 \\
0.417--0.450 &0.4325 & 0.345 &0.008 &0.009 & 0.433 & 0.012 & 0.014 \\
0.450--0.482 &0.4651 & 0.368 &0.010 &0.011 & 0.351 & 0.011 & 0.012 \\
0.482--0.526 &0.5035 & 0.384 &0.011 &0.013 & 0.264 & 0.008 & 0.010 \\
0.526--0.570 &0.5470 & 0.411 &0.016 &0.015 & 0.188 & 0.008 & 0.008 \\
0.570--0.658 &0.6083 & 0.439 &0.017 &0.017 & 0.122 & 0.005 & 0.006 \\
0.658--0.768 &0.7047 & 0.431 &0.032 &0.018 &0.0485 &0.0037 &0.0027 \\
0.768--1.000 &0.8383 & 0.328 &0.090 &0.042 &0.0078 &0.0022 &0.0011
 \\[.1cm] \hline \hline 
& \multicolumn{4}{c|}{ } & \multicolumn{3}{c|}{  }   \\[-.3cm]
0.014--1.000 & \multicolumn{4}{c|}{ }& 2.074 & 0.006 & 0.066
 \\[.1cm] \hline 
 \end{tabular}
\caption{ \baselineskip=12pt  \label{kafraxa}
Charged kaon fraction and differential production rate per hadronic
$Z^0$ decay.}
\end{center}
\end{table}

\begin{table}
\begin{center}
 \begin{tabular}{|c|c|r@{$\pm$}r@{$\pm$}r|r@{$\pm$}l@{$\pm$}l|} \hline
 $x_p$ Range & $<\! x_p\! >$ & \multicolumn{3}{c|}{$f_{\rm p}$}  &
      \multicolumn{3}{c|}{(1/N$_{evts}$) d$n_{\rm p}$/d$x_p$} \\ \hline 
  & & \multicolumn{3}{c|}{ } & \multicolumn{3}{c|}{  }   \\[-.3cm]
0.014--0.016 &0.0153 & 0.036 &0.001 &0.013 & 14.51 &  0.52 &  5.08 \\
0.016--0.022 &0.0191 & 0.052 &0.001 &0.008 & 17.32 &  0.27 &  2.58 \\
0.022--0.027 &0.0246 & 0.052 &0.001 &0.009 & 13.75 &  0.29 &  2.50 \\
0.027--0.033 &0.0301 & 0.052 &0.001 &0.006 & 11.12 &  0.17 &  1.24 \\
0.033--0.038 &0.0356 & 0.060 &0.001 &0.003 & 10.75 &  0.14 &  0.47 \\
0.038--0.044 &0.0411 & 0.060 &0.001 &0.002 & 9.048 & 0.123 & 0.350 \\
0.044--0.049 &0.0465 & 0.060 &0.001 &0.002 & 7.669 & 0.117 & 0.298 \\
0.049--0.055 &0.0521 & 0.066 &0.001 &0.003 & 7.410 & 0.113 & 0.294 \\
0.055--0.060 &0.0576 & 0.068 &0.001 &0.003 & 6.587 & 0.109 & 0.259 \\
0.060--0.066 &0.0630 & 0.067 &0.001 &0.003 & 5.788 & 0.105 & 0.238 \\
0.066--0.071 &0.0685 & 0.070 &0.001 &0.003 & 5.344 & 0.100 & 0.228 \\
0.071--0.077 &0.0740 & 0.073 &0.002 &0.003 & 4.987 & 0.104 & 0.229 \\
0.077--0.082 &0.0795 & 0.069 &0.002 &0.004 & 4.278 & 0.100 & 0.242 \\
0.082--0.088 &0.0850 & 0.074 &0.002 &0.005 & 4.117 & 0.101 & 0.253 \\
0.088--0.099 &0.0931 & 0.076 &0.002 &0.006 & 3.633 & 0.072 & 0.269 \\
0.099--0.110 &0.1040 & 0.076 &0.002 &0.008 & 3.036 & 0.076 & 0.300 \\
0.110--0.121 &0.1150 & 0.077 &0.002 &0.011 & 2.568 & 0.081 & 0.357 \\
0.121--0.143 &0.1310 & 0.081 &0.003 &0.015 & 2.165 & 0.069 & 0.398 \\
0.143--0.164 &0.1530 & 0.097 &0.005 &0.023 & 1.931 & 0.096 & 0.452 \\
0.164--0.186 &0.1750 & 0.106 &0.009 &0.039 & 1.603 & 0.133 & 0.594 \\
0.186--0.208 &0.1970 & 0.074 &0.004 &0.022 & 0.871 & 0.045 & 0.255 \\
0.208--0.230 &0.2189 & 0.098 &0.003 &0.019 & 0.912 & 0.030 & 0.179 \\
0.230--0.252 &0.2410 & 0.104 &0.003 &0.008 & 0.775 & 0.025 & 0.062 \\
0.252--0.274 &0.2629 & 0.108 &0.004 &0.007 & 0.639 & 0.022 & 0.044 \\
0.274--0.296 &0.2849 & 0.106 &0.004 &0.007 & 0.511 & 0.019 & 0.033 \\
0.296--0.318 &0.3068 & 0.109 &0.004 &0.006 & 0.419 & 0.016 & 0.024 \\
0.318--0.351 &0.3338 & 0.120 &0.004 &0.006 & 0.358 & 0.011 & 0.018 \\
0.351--0.384 &0.3666 & 0.115 &0.004 &0.005 & 0.254 & 0.009 & 0.012 \\
0.384--0.417 &0.3997 & 0.105 &0.004 &0.005 & 0.173 & 0.008 & 0.008 \\
0.417--0.450 &0.4325 & 0.113 &0.005 &0.004 & 0.141 & 0.007 & 0.005 \\
0.450--0.482 &0.4651 & 0.100 &0.006 &0.003 &0.0950 &0.0055 &0.0036 \\
0.482--0.526 &0.5035 & 0.100 &0.005 &0.003 &0.0688 &0.0039 &0.0027 \\
0.526--0.570 &0.5470 & 0.103 &0.007 &0.003 &0.0470 &0.0032 &0.0018 \\
0.570--0.658 &0.6083 & 0.087 &0.006 &0.003 &0.0241 &0.0017 &0.0012 \\
0.658--0.768 &0.7047 & 0.083 &0.009 &0.004 &0.0093 &0.0010 &0.0006 \\
0.768--1.000 &0.8383 & 0.062 &0.014 &0.005 &0.0015 &0.0003 &0.0001
 \\[.1cm] \hline \hline 
& \multicolumn{4}{c|}{ } & \multicolumn{3}{c|}{  }   \\[-.3cm]
0.014--1.000 & \multicolumn{4}{c|}{  }  & 0.984 & 0.006 & 0.035
 \\[.1cm] \hline 
 \end{tabular}
\caption{ \baselineskip=12pt  \label{pfraxa}
Proton plus antiproton
fraction and differential production rate per hadronic $Z^0$ decay.}
\end{center}
\end{table}

\begin{figure}
 \vspace{-0.2cm}
  \epsfxsize=6.5in
  \begin{center}\mbox{\epsffile{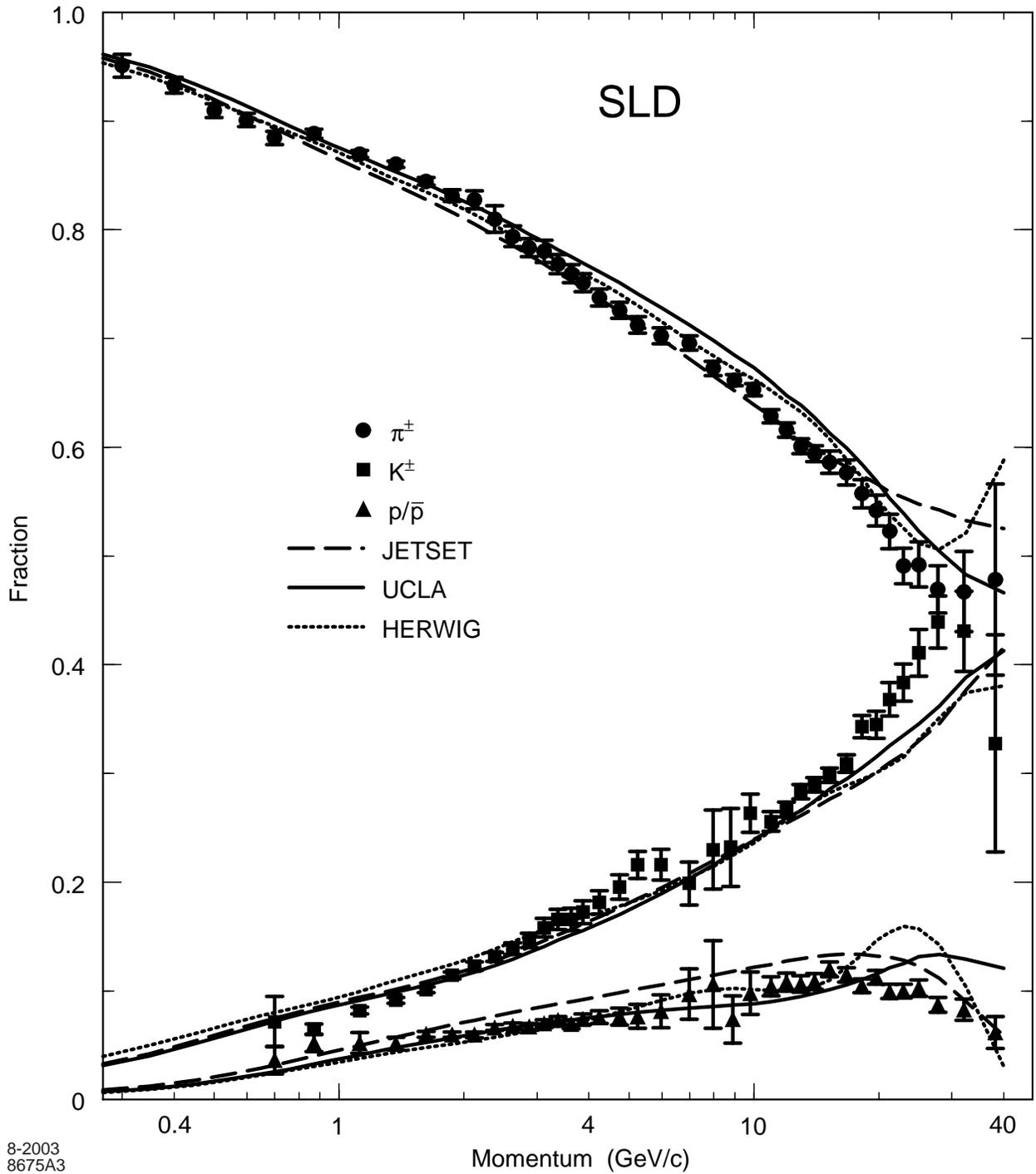}}\end{center}
 \vspace{-0.8cm}
 \caption{ \baselineskip=12pt  \label{fallmc}
Comparison of our measured charged hadron fractions (symbols)
with the predictions of the JETSET (dashed lines), UCLA (solid lines) and HERWIG
(dotted lines) models.
    }
\end{figure} 

\begin{figure}
 \vspace{-1.cm}
  \epsfxsize=6.5in
  \begin{center}\mbox{\epsffile{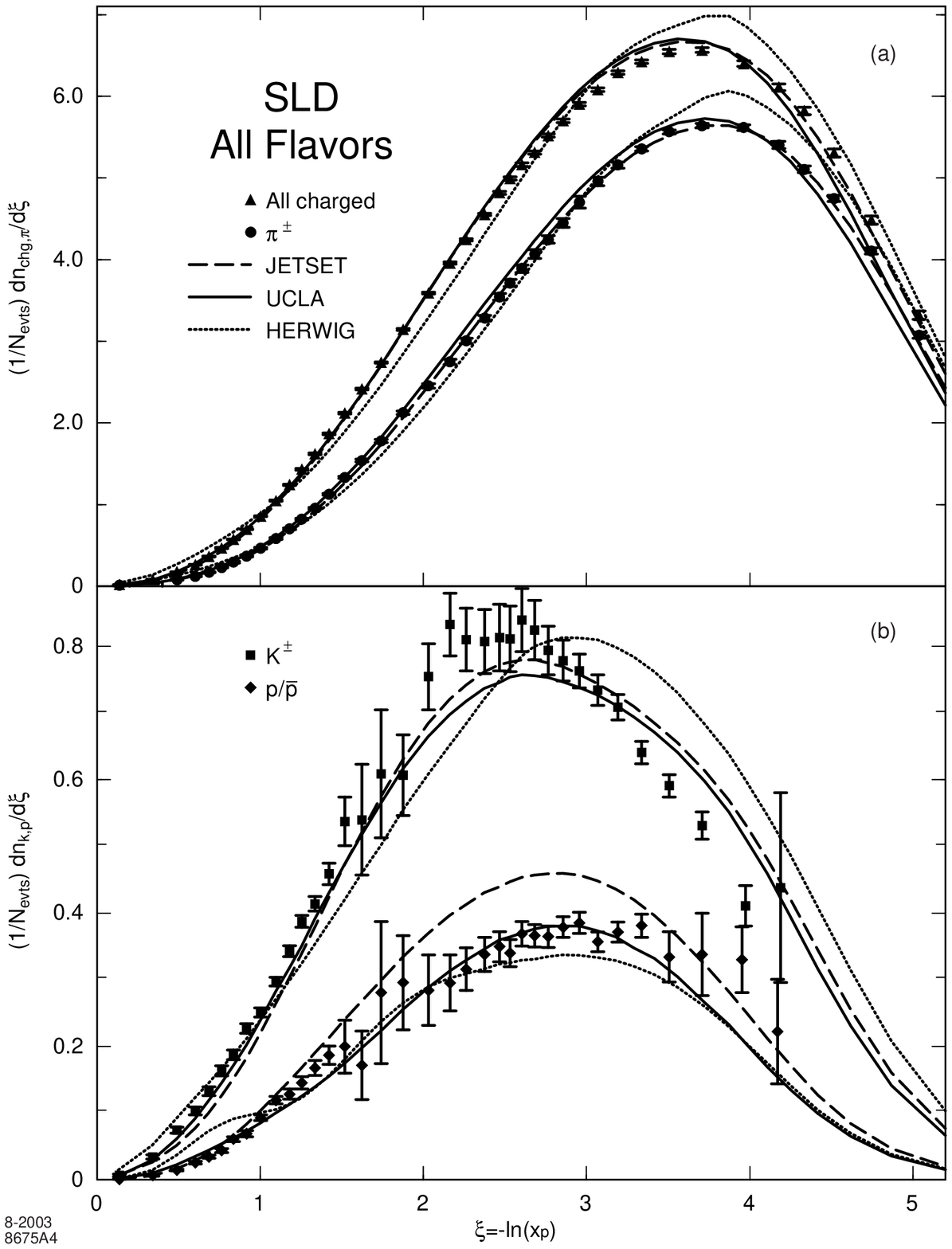}}\end{center}
 \vspace{-0.8cm}
 \caption{ \baselineskip=12pt  \label{xsallmc}
Comparison of our measured (a) charged particle and pion and (b) kaon and proton
production rates (symbols)
with the predictions of the JETSET (dashed lines), UCLA (solid lines) and HERWIG
(dotted lines) models.
    }
\end{figure} 

Pions are seen to dominate the charged hadron production at low momentum,
and to decline steadily in fraction as momentum increases.
The kaon fraction rises steadily, approaching the pion fraction at high
momentum.
The proton fraction rises to a plateau value of about one-tenth at about
10 GeV/c, then declines at the highest momenta.

In fig.~\ref{fallmc} we compare our results with the predictions of the
JETSET, UCLA and HERWIG models.
The momentum dependence for each of the three hadron species 
is reproduced qualitatively by all models.
All three models overestimate the kaon fraction for momenta below about
1.5 GeV/c, and underestimate it for momenta above about 15 GeV/c.
The UCLA model overestimates the pion fraction by about 2$\sigma$
(taking into account the correlation in the experimental errors)
in the 5--15 GeV/c range, but is the only model consistent with the behavior
above about 20 GeV/c.
The JETSET model overestimates the proton fraction at all momenta, but describes
the momentum dependence;
HERWIG and UCLA predict behavior at high momentum that is inconsistent with
the proton data.

Since the JETSET model has a number of parameters that control specific aspects
of relative particle production, we have investigated some simple changes.
We find that reducing the diquark:quark ratio (PARJ(1)) from 0.10 to 0.08, while
leaving all other parameters unchanged,
results in a good description of the proton fraction.
The kaon fraction is sensitive to both the $s$-quark probability (PARJ(2)) and
the strange vector:pseudoscalar ratio (PARJ(12)), however no combination of
these two parameters was found to give a good description of the kaon fraction
over the entire measured momentum range.

In order to obtain charged hadron production rates, the simulation
was used to subtract
the contribution of all particles (mostly leptons) that were not charged pions,
kaons or protons from our measured total charged production rate
(table~\ref{chgxs}).
The pion, kaon and proton fractions were multiplied by this adjusted rate to
obtain the individual rates tabulated in
tables~\ref{pifraxa}--\ref{pfraxa}
as a function of scaled momentum.
In fig.~\ref{xsallmc} we compare our $\xi$ distributions with the predictions of
the three models.
The features for pions are similar to those for all charged tracks:
all models describe the data qualitatively;
JETSET is within about 2\% except for $\xi<1.5$ and $\xi>4.5$;
UCLA shows a spectrum very similar to the data but shifted slightly toward
lower $\xi$ values;
HERWIG is several percent high (low) for $\xi<1$ and $3.3<\xi<4.3$
($1.5<\xi<2.5$).
For the protons we see features similar to those seen in the proton fraction
(fig.~\ref{fallmc}):
JETSET describes the shape of the spectrum but is high by about 20\%;
UCLA describes the data well except for the structure for $\xi<0.8$;
HERWIG is consistent with the data for $\xi>1.5$, but shows a pronounced
structure at lower values of $\xi$ that is inconsistent with the data.
For kaons, all three predictions are too high for $\xi>3.3$, especially HERWIG,
which is also quite high for $\xi<0.6$ and low for $1.2<\xi<2.4$;
the other two predictions are low for $\xi<0.8$, and consistent with the data
elsewhere, although they peak at higher $\xi$ values than the data.

Our fractions and production rates are generally consistent with those
from previous experiments at the $Z^0$~\cite{chgxsa,chgxsd,opalfr}.
The ALEPH pion spectrum lies above ours at large $\xi$;
the OPAL proton spectrum is lower than all others at small $\xi$;
all other differences are within two standard deviations, considering
correlations in the systematic errors.
Our measurement is the most precise in several regions, most notably for pions
in the range $3<\xi<4$, kaons in the range $2.5<\xi<4$ and protons in the range
$2.5<\xi<3.5$.
Measurements based on ring imaging and those based on ionization energy loss
rates cover complementary momentum ranges and could be combined to provide
continuous, high-precision coverage over the range from the beam momentum
($\xi=0$) down to $p=0.22$ GeV/c ($\xi=5.3$).

\section{Flavor-Dependent Analysis}

The analyses described above were repeated on the light-, $c$-, $b$-, and
untagged event samples described in section 3, to yield differential production
rates $R^{ktag}_h$, $k=l$, $c$, $b$, $un$, respectively;
the correction for leptons was not performed at this point, so that
$h=e\mu\pi$, $K$, p.
True differential rates $R^m_h$ in events of the three flavor types,
$m=l$, $c$, $b$, representing events of the types
$Z^0 \rightarrow u\bar{u},d\bar{d},s\bar{s}$,
$Z^0 \rightarrow c\bar{c}$, and $Z^0 \rightarrow b\bar{b}$,
respectively, were extracted by solving for each species
$h$ the relations:
\begin{equation}
R^{ktag}_h = \frac{\Sigma_m B^h_{mk} \epsilon_{mk} F_m R^m_h}
                   {\Sigma_m          \epsilon_{mk} F_m} .
\end{equation}
Here, $F_m$ is the fraction of hadronic $Z^0$ decays of flavor type $m$, taken
from the Standard Model, $\epsilon_{mk}$ is an element of the event tagging
efficiency matrix (see table~\ref{tlveff}), 
and $B_{mk}^h$ represents the momentum-dependent bias of tag $k$ toward
selecting events of flavor $m$ that contain hadrons of species $h$.
Ideally all biases would be unity in this formulation.
The biases were calculated from the simulation as
$B^h_{mk}=(n^h_{m,ktag}/N_{m,ktag})/(n^h_m/N_m)$, where $N_m$ $(n^h_m)$
is the number of simulated events (hadrons of species $h$ in events) of true
flavor $m$ and $N_{m,ktag}$ $(n^h_{m,ktag})$ is the number of ($h$-hadrons in)
those events that are tagged as flavor $k$.
The diagonal bias values \cite{hjkang} are within a few percent of unity
reflecting the small dependences of the flavor tags on the charged multiplicity
of the event and tracks from strange mesons and baryons that decay close to the
IP.
The sum of the products of biases and efficiencies must be unity for
events of a given true flavor $m$, $\Sigma_k B^h_{mk} \epsilon_{mk}=1$;
some off-diagonal bias values therefore deviate substantially from unity since
the corresponding mistagging rate can be small, but these do not affect the
results as the figure of merit is the diagonal element.

The unfolded pion rates were corrected for the contributions
from leptons by subtracting the absolute lepton rate predicted by the
simulation.
At low momentum this background is dominated by electrons from photon
conversions, and is a 6\% contribution at 1 GeV/c that falls rapidly with
increasing momentum.
For the heavy flavors, electrons and muons from semi-leptonic decays
of heavy hadrons cause the correction to increase with momentum above about 5
GeV/c, reaching 13\% (60\%) for $c$-($b$-)flavor events in the highest momentum
bins.

The resulting differential production rates are listed
in tables \ref{xsfpi}--\ref{xsfpr}.
The systematic errors listed are only those relevant for the comparison of
different flavors, namely those due to uncertainties in the unfolding procedure;
the relative systematic errors given in the preceding section are also
applicable, but are common to all three flavor categories;
we also list ratios, for which these common errors cancel.
The flavor unfolding systematic errors were evaluated by varying:
each diagonal element of the event tagging efficiency matrix $\epsilon_{ii}$
by $\pm$0.01;
the heavy quark production fractions $R_b$ and $R_c$ by the errors
on their respective world averages;
each diagonal bias value $B_{ii}^h$ by the larger of $\pm$0.005 and $\pm$20\%
of its difference from unity;
the photon conversion rate in the simulation by $\pm$15\%;
and the simulated physics lepton rates in light-, $c$-, and $b$-flavor
events by $\pm$20\%, $\pm$10\% and $\pm$5\%, respectively.
These variations correspond to uncertainties on measurements from our data or
other experiments except for the variation of the bias, which was chosen
conservatively to be larger than any change seen in a set of comparisons of
relevant quantities in data and simulation when selection cuts were varied.
The unfolding systematic errors are similar in magnitude to (smaller than) the
statistical errors at low (high) momenta, and are generally dominated by the
bias for the relevant flavor.
There are also substantial contributions from the lepton correction at high
momentum for $R^{\pi}_c$ and $R^{\pi}_b$, and from all three bias terms for
$R^{K}_c$ and $R^{\rm p}_c$.

\begin{table}
\begin{center}
 \begin{tabular}{|c|r@{$\pm$}l|r@{$\pm$}l|
                    r@{$\pm$}l|r@{$\pm$}l|r@{$\pm$}l|} \hline
& \multicolumn{6}{|c|}{ } & \multicolumn{4}{|c|}{ } \\ [-.4cm]
& \multicolumn{6}{|c|}{Pion Production Rates} & \multicolumn{4}{|c|}{Ratios} \\
  $<\!\!x_p\!\!>$
& \multicolumn{2}{|c|}{$u\bar{u}$, $d\bar{d}$, $s\bar{s}$}
& \multicolumn{2}{|c|}{$c\bar{c}$}
& \multicolumn{2}{|c|}{$b\bar{b}$}
& \multicolumn{2}{|c|}{$c$:$uds$}
& \multicolumn{2}{|c|}{$b$:$uds$} \\[.05cm]\hline
0.0066 &474.0& 13.9&425.5& 26.6&478.1& 15.8& 0.898&0.058&1.009&0.019 \\[-0.05cm]
0.0088 &467.3& 10.5&440.5& 23.2&488.4& 11.9& 0.943&0.054&1.045&0.017 \\[-0.05cm]
0.0109 &418.2&  8.4&453.8& 20.0&463.7&  9.5& 1.085&0.055&1.109&0.018 \\[-0.05cm]
0.0131 &375.5&  6.9&409.2& 17.2&432.2&  7.7& 1.090&0.053&1.151&0.019 \\[-0.05cm]
0.0153 &327.7&  5.7&372.8& 14.6&382.4&  6.5& 1.138&0.052&1.167&0.019 \\[-0.05cm]
0.0191 &275.8&  4.2&306.4& 10.9&333.3&  4.6& 1.111&0.047&1.208&0.017 \\[-0.05cm]
0.0246 &216.0&  3.0&234.6&  8.0&261.7&  3.3& 1.086&0.044&1.212&0.018 \\[-0.05cm]
0.0301 &171.2&  2.2&197.4&  6.2&214.2&  2.7& 1.153&0.043&1.251&0.017 \\[-0.05cm]
0.0356 &140.4&  1.9&155.8&  6.0&175.2&  2.3& 1.109&0.050&1.247&0.016 \\[-0.05cm]
0.0411 &116.4&  1.5&132.5&  4.3&145.4&  1.9& 1.138&0.043&1.249&0.018 \\[-0.05cm]
0.0465 & 99.9&  1.2&109.3&  3.5&121.4&  1.7& 1.095&0.040&1.215&0.018 \\[-0.05cm]
0.0521 & 85.4&  1.0& 92.9&  2.9&103.3&  1.5& 1.088&0.038&1.209&0.019 \\[-0.05cm]
0.0576 &72.85& 0.89&77.56& 2.48&89.24& 1.36& 1.065&0.037&1.225&0.020 \\[-0.05cm]
0.0630 &64.51& 0.79&68.23& 2.17&75.47& 1.21& 1.058&0.036&1.170&0.020 \\[-0.05cm]
0.0685 &56.82& 0.72&60.06& 1.97&65.97& 1.12& 1.057&0.038&1.161&0.022 \\[-0.05cm]
0.0740 &50.84& 0.66&51.76& 1.81&59.39& 1.04& 1.018&0.038&1.168&0.023 \\[-0.05cm]
0.0795 &45.34& 0.61&45.28& 1.67&52.11& 0.97& 0.999&0.039&1.149&0.024 \\[-0.05cm]
0.0850 &40.71& 0.56&40.04& 1.55&45.86& 0.90& 0.984&0.041&1.127&0.025 \\[-0.05cm]
0.0931 &34.60& 0.40&33.50& 1.12&38.29& 0.65& 0.968&0.034&1.106&0.020 \\[-0.05cm]
0.1040 &28.99& 0.35&27.45& 0.99&30.57& 0.58& 0.947&0.036&1.054&0.021 \\[-0.05cm]
0.1150 &24.19& 0.31&22.92& 0.87&24.34& 0.51& 0.947&0.038&1.006&0.022 \\[-0.05cm]
0.1310 &18.97& 0.22&18.73& 0.63&18.21& 0.36& 0.987&0.036&0.960&0.019 \\[-0.05cm]
0.1530 &14.52& 0.17&13.72& 0.50&12.27& 0.28& 0.945&0.038&0.845&0.020 \\[-0.05cm]
0.1750 &11.06& 0.14&10.18& 0.41& 8.25& 0.22& 0.920&0.040&0.746&0.021 \\[-0.05cm]
0.1970 & 8.67& 0.12& 7.53& 0.34& 5.83& 0.18& 0.868&0.042&0.673&0.022 \\[-0.05cm]
0.2189 & 6.79& 0.10& 5.76& 0.29& 4.14& 0.15& 0.848&0.046&0.609&0.023 \\[-0.05cm]
0.2410 &5.341&0.085&4.381&0.235&2.984&0.128& 0.820&0.048&0.559&0.025 \\[-0.05cm]
0.2629 &4.214&0.073&3.358&0.202&2.303&0.110& 0.797&0.051&0.547&0.027 \\[-0.05cm]
0.2849 &3.452&0.064&2.487&0.171&1.642&0.094& 0.720&0.052&0.475&0.028 \\[-0.05cm]
0.3068 &2.727&0.056&1.947&0.148&1.365&0.085& 0.714&0.057&0.501&0.032 \\[-0.05cm]
0.3338 &2.138&0.042&1.436&0.108&0.886&0.063& 0.672&0.053&0.414&0.030 \\[-0.05cm]
0.3666 &1.652&0.036&0.817&0.087&0.631&0.052& 0.495&0.054&0.382&0.032 \\[-0.05cm]
0.3997 &1.164&0.031&0.614&0.074&0.490&0.047& 0.527&0.066&0.421&0.042 \\[-0.05cm]
0.4325 &0.874&0.027&0.386&0.063&0.276&0.038& 0.441&0.074&0.316&0.045 \\[-0.05cm]
0.4651 &0.622&0.024&0.429&0.061&0.187&0.033& 0.689&0.102&0.301&0.054 \\[-0.05cm]
0.5035 &0.441&0.019&0.206&0.043&0.111&0.025& 0.467&0.101&0.252&0.058 \\[-0.05cm]
0.5470 &0.300&0.017&0.142&0.037&0.045&0.019& 0.472&0.128&0.151&0.062 \\[-0.05cm]
0.6083 &0.178&0.010&0.066&0.021&0.039&0.010& 0.371&0.122&0.222&0.060 \\[-0.05cm]
0.7047 &0.081&0.007&0.003&0.010&0.011&0.005& 0.043&0.125&0.141&0.062 \\[-0.05cm]
0.8383 &0.016&0.003&0.003&0.006&0.003&0.002& 0.160&0.356&0.190&0.150
 \\[.05cm] \hline \hline 
& \multicolumn{2}{|c|}{ } & \multicolumn{2}{|c|}{ } & \multicolumn{2}{|c|}{ } &
  \multicolumn{4}{|c|}{ } \\ [-.4cm]
Total & 15.294 & 0.250 & 15.783 & 0.465  & 16.841 & 0.278 &
  \multicolumn{4}{|c|}{ } \\
  [.05cm] \hline
  \end{tabular}
\caption{\baselineskip=12pt  \label{xsfpi}
Differential production rates (1/N$_{evts}$)d$n_{\pi^{\pm}}$/d$x_p$ of
charged pions per $Z^0$ decay into light ($u$, $d$, $s$), $c$ and
$b$ primary flavors.
The errors are the sum in quadrature of statistical errors and those systematic
uncertainties arising from the unfolding procedure;  systematic errors common
to the three flavors are not included in the rates and cancel in the ratios.
The $<\!\! x_p \!\!>$ values for the three
flavor samples are consistent, and have been averaged.}
\end{center}
\end{table}

\begin{table}
\begin{center}
 \begin{tabular}{|c|r@{$\pm$}l|r@{$\pm$}l|
                    r@{$\pm$}l|r@{$\pm$}l|r@{$\pm$}l|} \hline
& \multicolumn{6}{|c|}{ } & \multicolumn{4}{|c|}{ } \\ [-.3cm]
& \multicolumn{6}{|c|}{Kaon Production Rates} &
  \multicolumn{4}{|c|}{Ratios} \\
  $<\!\!x_p\!\!>$
& \multicolumn{2}{|c|}{$u\bar{u}$, $d\bar{d}$, $s\bar{s}$}
& \multicolumn{2}{|c|}{$c\bar{c}$}
& \multicolumn{2}{|c|}{$b\bar{b}$}
& \multicolumn{2}{|c|}{$c$:$uds$}
& \multicolumn{2}{|c|}{$b$:$uds$} \\[.1cm]\hline
0.0153 &27.05& 1.27 & 30.92& 3.86 & 30.21& 1.99 &  1.143&0.155 & 1.117&0.090 \\
0.0191 &20.00& 0.42 & 22.43& 1.37 & 23.06& 0.62 &  1.121&0.078 & 1.153&0.040 \\
0.0246 &19.74& 0.40 & 22.04& 1.27 & 22.89& 0.60 &  1.116&0.074 & 1.159&0.039 \\
0.0301 &17.52& 0.37 & 20.82& 1.17 & 21.64& 0.55 &  1.189&0.076 & 1.236&0.041 \\
0.0356 &16.08& 0.37 & 16.79& 1.15 & 21.36& 0.53 &  1.044&0.083 & 1.328&0.046 \\
0.0411 &15.04& 0.34 & 16.68& 1.15 & 21.36& 0.56 &  1.110&0.086 & 1.420&0.048 \\
0.0465 &13.54& 0.34 & 16.46& 1.06 & 19.90& 0.57 &  1.215&0.088 & 1.469&0.056 \\
0.0521 &11.87& 0.34 & 15.81& 1.08 & 18.91& 0.60 &  1.332&0.103 & 1.593&0.068 \\
0.0576 &11.44& 0.33 & 12.62& 0.99 & 18.46& 0.58 &  1.103&0.094 & 1.613&0.069 \\
0.0630 &10.64& 0.30 & 12.24& 0.92 & 17.43& 0.54 &  1.151&0.095 & 1.639&0.069 \\
0.0685 &10.24& 0.29 & 11.42& 0.87 & 16.92& 0.53 &  1.115&0.093 & 1.652&0.070 \\
0.0740 & 9.67& 0.29 & 10.95& 0.85 & 15.62& 0.52 &  1.133&0.096 & 1.616&0.072 \\
0.0795 & 8.13& 0.27 & 10.88& 0.84 & 15.11& 0.52 &  1.339&0.114 & 1.859&0.090 \\
0.0850 & 7.98& 0.28 &  9.62& 0.81 & 13.18& 0.50 &  1.206&0.111 & 1.651&0.084 \\
0.0931 & 7.00& 0.19 &  9.84& 0.59 & 12.43& 0.36 &  1.406&0.094 & 1.778&0.070 \\
0.1040 & 6.36& 0.19 &  8.08& 0.58 & 11.56& 0.37 &  1.271&0.100 & 1.819&0.080 \\
0.1150 & 5.85& 0.20 &  8.98& 0.63 &  9.96& 0.38 &  1.535&0.122 & 1.704&0.088 \\
0.1310 & 4.89& 0.15 &  6.59& 0.45 &  7.17& 0.27 &  1.349&0.104 & 1.467&0.072 \\
0.1530 & 3.41& 0.17 &  5.50& 0.51 &  4.58& 0.29 &  1.614&0.175 & 1.343&0.108 \\
0.1750 & 2.84& 0.22 &  5.12& 0.68 &  4.20& 0.36 &  1.805&0.289 & 1.480&0.173 \\
0.1970 &2.564&0.082 & 3.850&0.245 & 2.541&0.126 &  1.502&0.110 & 0.991&0.059 \\
0.2189 &2.401&0.067 & 3.087&0.190 & 2.009&0.096 &  1.286&0.090 & 0.837&0.047 \\
0.2410 &1.973&0.054 & 2.074&0.145 & 1.627&0.078 &  1.051&0.081 & 0.825&0.046 \\
0.2629 &1.643&0.048 & 1.960&0.132 & 1.116&0.062 &  1.193&0.090 & 0.679&0.043 \\
0.2849 &1.481&0.044 & 1.681&0.119 & 0.830&0.053 &  1.135&0.090 & 0.560&0.039 \\
0.3068 &1.211&0.039 & 1.368&0.104 & 0.640&0.045 &  1.129&0.096 & 0.529&0.041 \\
0.3338 &1.001&0.029 & 1.043&0.076 & 0.452&0.032 &  1.042&0.084 & 0.451&0.034 \\
0.3666 &0.746&0.025 & 0.874&0.068 & 0.337&0.028 &  1.171&0.102 & 0.451&0.040 \\
0.3997 &0.666&0.023 & 0.600&0.058 & 0.245&0.024 &  0.900&0.094 & 0.367&0.038 \\
0.4325 &0.559&0.022 & 0.408&0.050 & 0.149&0.020 &  0.730&0.094 & 0.266&0.036 \\
0.4651 &0.426&0.020 & 0.408&0.050 & 0.108&0.018 &  0.957&0.126 & 0.253&0.043 \\
0.5035 &0.363&0.016 & 0.243&0.037 & 0.057&0.012 &  0.669&0.106 & 0.156&0.035 \\
0.5470 &0.261&0.015 & 0.173&0.034 & 0.061&0.013 &  0.663&0.136 & 0.233&0.050 \\
0.6083 &0.183&0.010 & 0.064&0.020 & 0.012&0.005 &  0.351&0.109 & 0.066&0.030 \\
0.7047 &0.079&0.007 & 0.009&0.011 & 0.002&0.003 &  0.113&0.141 & 0.026&0.035 \\
0.8383 &0.008&0.004 & 0.008&0.008 &--.001&0.001 &  0.989&1.126 &--.098&0.102
 \\[.1cm] \hline \hline 
& \multicolumn{2}{|c|}{ } & \multicolumn{2}{|c|}{ } & \multicolumn{2}{|c|}{ } &
  \multicolumn{4}{|c|}{ } \\ [-.3cm]
Total & 1.869 & 0.062  & 2.273 & 0.093  &  2.377 & 0.080 &
  \multicolumn{4}{|c|}{ } \\
  [.1cm] \hline
\end{tabular}
\caption{\baselineskip=12pt  \label{xsfka}
Differential production rates (1/N$_{evts}$)d$n_{K^{\pm}}$/d$x_p$
of $K^\pm$ mesons per $Z^0$ decay into light, $c$ and $b$ primary flavors.
}
\end{center}
\end{table}

\begin{table}
\begin{center}
 \begin{tabular}{|c|r@{$\pm$}l|r@{$\pm$}l|
                    r@{$\pm$}l|r@{$\pm$}l|r@{$\pm$}l|} \hline
& \multicolumn{6}{|c|}{ } & \multicolumn{4}{|c|}{ } \\ [-.3cm]
& \multicolumn{6}{|c|}{Proton Production Rates} & \multicolumn{4}{|c|}{Ratios} \\
$<\!\!x_p\!\!>$
& \multicolumn{2}{|c|}{$u\bar{u}$, $d\bar{d}$, $s\bar{s}$}
& \multicolumn{2}{|c|}{$c\bar{c}$}
& \multicolumn{2}{|c|}{$b\bar{b}$}
& \multicolumn{2}{|c|}{$c$:$uds$}
& \multicolumn{2}{|c|}{$b$:$uds$} \\[.1cm]\hline
0.0153 &13.98& 0.99&13.28& 2.94& 13.79& 1.49 & 0.950&0.226 & 0.987&0.127 \\
0.0191 &17.63& 0.58&15.22& 1.89& 17.93& 0.78 & 0.863&0.120 & 1.017&0.056 \\
0.0246 &13.42& 0.60&13.32& 1.86& 16.41& 0.93 & 0.992&0.152 & 1.223&0.089 \\
0.0301 &10.57& 0.36& 9.60& 1.16& 12.11& 0.52 & 0.909&0.122 & 1.146&0.063 \\
0.0356 & 9.98& 0.31&11.64& 1.01& 10.32& 0.42 & 1.166&0.118 & 1.034&0.052 \\
0.0411 & 8.37& 0.26&10.07& 0.87&  9.52& 0.40 & 1.203&0.118 & 1.138&0.059 \\
0.0465 & 7.33& 0.24& 8.10& 0.76&  7.72& 0.36 & 1.105&0.117 & 1.054&0.059 \\
0.0521 & 7.79& 0.23& 6.09& 0.72&  6.86& 0.32 & 0.781&0.099 & 0.879&0.049 \\
0.0576 & 6.62& 0.22& 6.54& 0.67&  6.19& 0.32 & 0.988&0.111 & 0.935&0.057 \\
0.0630 & 5.88& 0.20& 6.36& 0.63&  4.96& 0.29 & 1.082&0.117 & 0.845&0.056 \\
0.0685 & 5.39& 0.19& 4.62& 0.59&  4.82& 0.29 & 0.857&0.115 & 0.895&0.062 \\
0.0740 & 5.22& 0.19& 4.43& 0.58&  4.57& 0.29 & 0.848&0.117 & 0.875&0.064 \\
0.0795 & 4.42& 0.18& 4.08& 0.55&  4.07& 0.29 & 0.924&0.133 & 0.920&0.076 \\
0.0850 & 4.44& 0.19& 3.67& 0.55&  3.82& 0.29 & 0.827&0.131 & 0.860&0.074 \\
0.0931 & 3.65& 0.13& 4.07& 0.42&  3.29& 0.21 & 1.116&0.123 & 0.901&0.066 \\
0.1040 & 3.11& 0.13& 2.98& 0.41&  2.68& 0.22 & 0.960&0.139 & 0.861&0.079 \\
0.1150 & 2.73& 0.15& 2.30& 0.43&  2.24& 0.23 & 0.844&0.166 & 0.819&0.096 \\
0.1310 & 2.15& 0.12& 2.39& 0.36&  1.84& 0.19 & 1.110&0.180 & 0.856&0.100 \\
0.1530 & 1.83& 0.16& 1.72& 0.50&  1.91& 0.27 & 0.940&0.287 & 1.048&0.174 \\
0.1750 & 1.84& 0.24& 0.31& 0.71&  1.25& 0.36 & 0.171&0.392 & 0.683&0.219 \\
0.1970 &0.905&0.078&0.561&0.235& 0.867&0.121 & 0.619&0.265 & 0.958&0.157 \\
0.2189 &1.065&0.054&0.978&0.163& 0.739&0.072 & 0.918&0.161 & 0.695&0.077 \\
0.2410 &0.822&0.044&0.907&0.136& 0.645&0.060 & 1.104&0.176 & 0.784&0.085 \\
0.2629 &0.762&0.038&0.652&0.116& 0.392&0.047 & 0.855&0.159 & 0.515&0.067 \\
0.2849 &0.628&0.033&0.572&0.101& 0.252&0.038 & 0.911&0.169 & 0.401&0.064 \\
0.3068 &0.486&0.029&0.494&0.089& 0.266&0.035 & 1.016&0.195 & 0.547&0.079 \\
0.3338 &0.446&0.022&0.454&0.069& 0.146&0.022 & 1.016&0.165 & 0.327&0.052 \\
0.3666 &0.306&0.018&0.314&0.054& 0.102&0.018 & 1.026&0.190 & 0.333&0.061 \\
0.3997 &0.230&0.015&0.170&0.041& 0.020&0.012 & 0.741&0.186 & 0.085&0.053 \\
0.4325 &0.197&0.013&0.103&0.033& 0.034&0.011 & 0.522&0.171 & 0.171&0.055 \\
0.4651 &0.145&0.011&0.064&0.028&--.004&0.007 & 0.445&0.196 &--.026&0.048 \\
0.5035 &0.108&0.008&0.015&0.018& 0.016&0.006 & 0.142&0.166 & 0.150&0.058 \\
0.5470 &0.070&0.006&0.044&0.017&--.003&0.003 & 0.626&0.243 &--.043&0.049 \\
0.6083 &0.036&0.003&0.007&0.007& 0.004&0.002 & 0.186&0.193 & 0.107&0.061 \\
0.7047 &0.013&0.002&0.015&0.005&--.001&0.001 & 1.150&0.434 &--.089&0.045 \\
0.8383 &0.003&0.001&0.001&0.002& 0.000&0.000 & 0.221&0.620 & 0.159&0.160
 \\[.1cm] \hline \hline 
& \multicolumn{2}{|c|}{ } & \multicolumn{2}{|c|}{ } & \multicolumn{2}{|c|}{ } &
  \multicolumn{4}{|c|}{ } \\ [-.3cm]
Total & 1.008 & 0.038  & 0.930 & 0.056  &  0.909 & 0.037 &
  \multicolumn{4}{|c|}{ } \\
  [.1cm] \hline
  \end{tabular}
\caption{\baselineskip=12pt  \label{xsfpr}
Differential production rates (1/N$_{evts}$)d$n_{\rm p/\bar{p}}$/d$x_p$
of p and $\bar{\rm p}$ per $Z^0$ decay into light, $c$ and $b$ primary flavors.
}
\end{center}
\end{table}

\begin{figure}
 \vspace{-0.2cm}
  \epsfxsize=6.5in
  \begin{center}\mbox{\epsffile{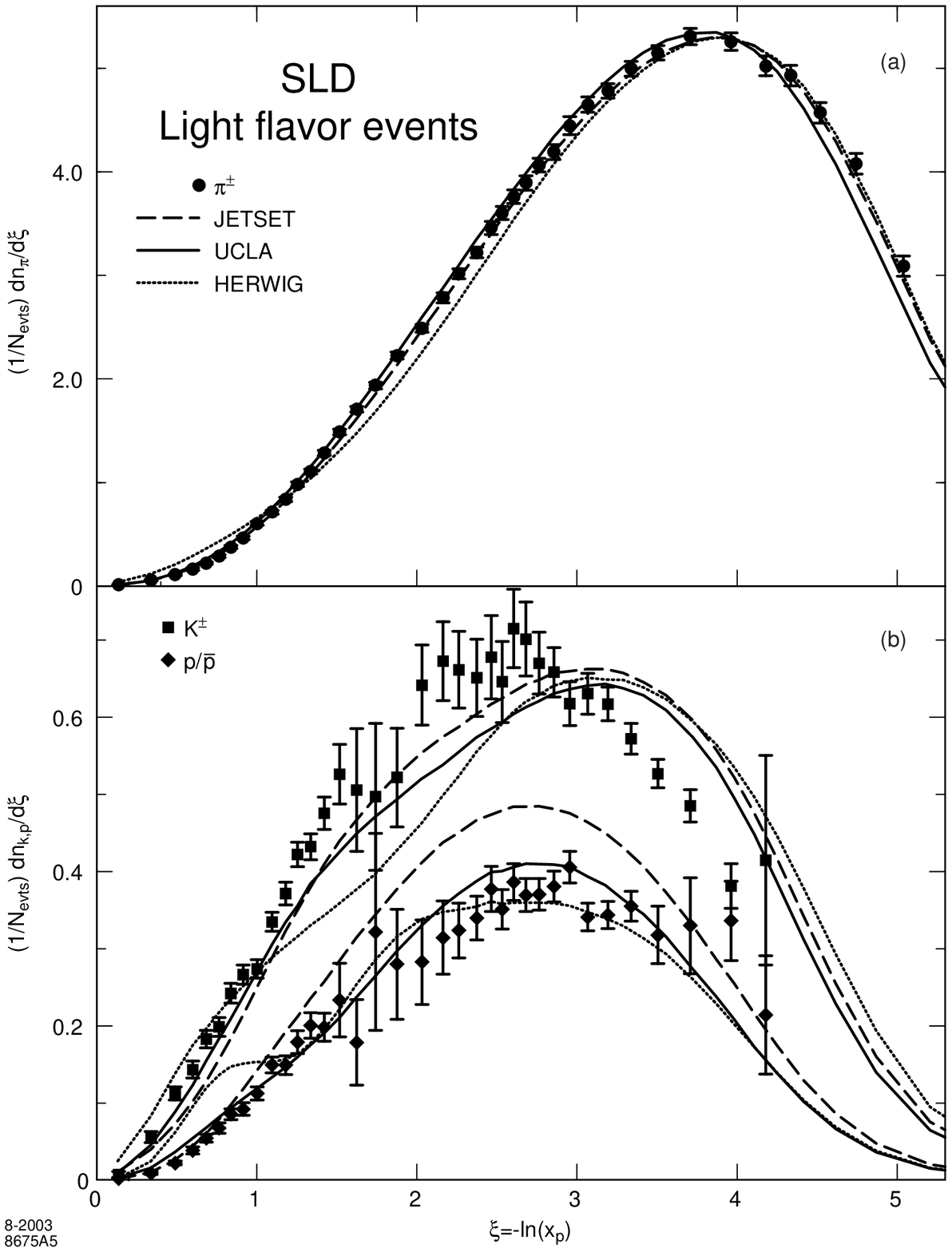}}\end{center}
 \vspace{-0.8cm}
 \caption{ \baselineskip=12pt  \label{xsudsmc}
Distributions of $\xi$ in light-flavor events for (a) pions (circles) and (b)
kaons (squares) and protons (diamonds), compared with the predictions of the
three models.
    }
\end{figure}

\begin{figure}
 \vspace{-0.2cm}
  \epsfxsize=6.5in
  \begin{center}\mbox{\epsffile{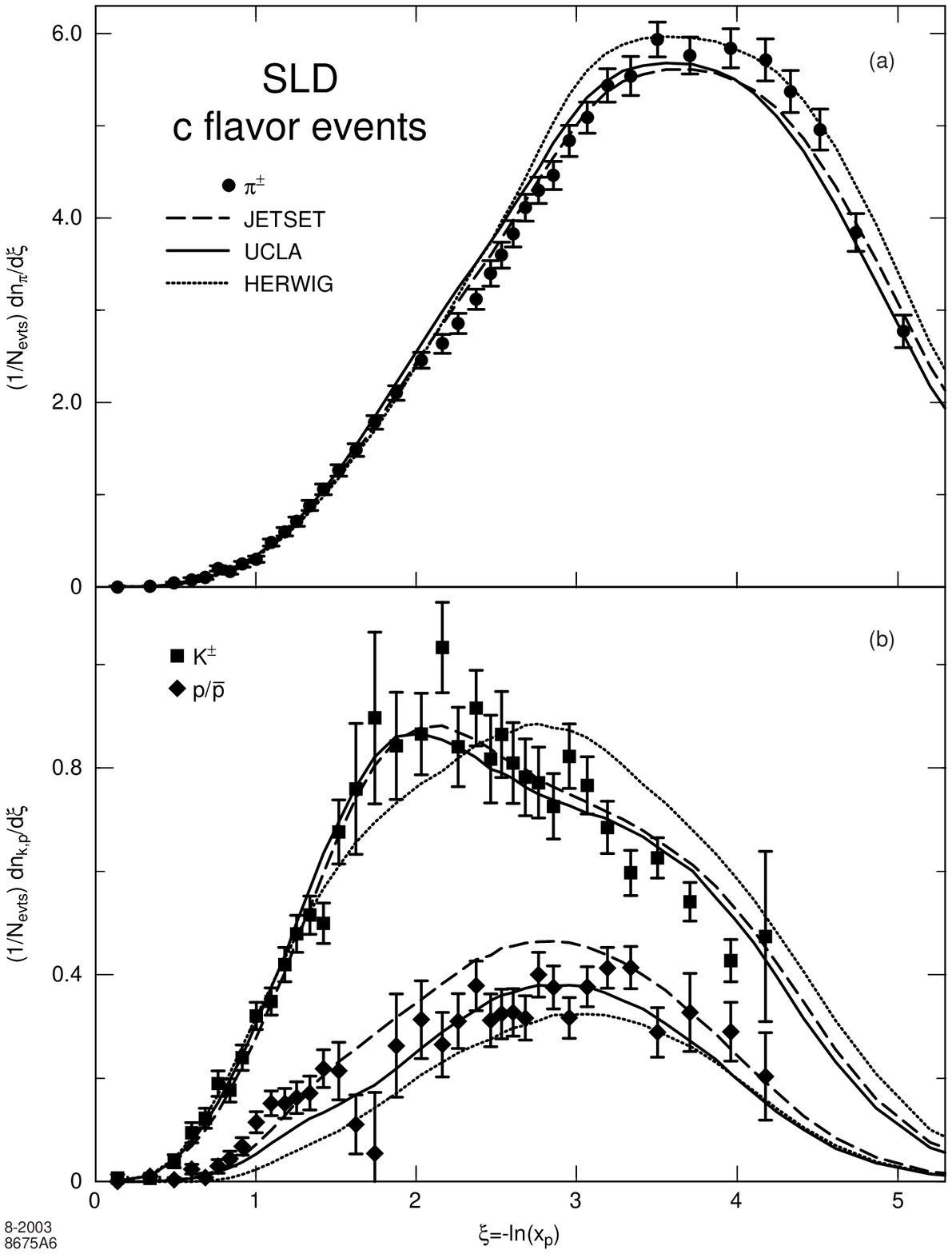}}\end{center}
 \vspace{-0.8cm}
 \caption{ \baselineskip=12pt  \label{xsccmc}
Distributions of $\xi$ in $c$-flavor events for (a) pions (circles) and (b)
kaons (squares) and protons (diamonds), compared with the predictions of the
three models.
    }
\end{figure}

\begin{figure}
 \vspace{-0.2cm}
  \epsfxsize=6.5in
  \begin{center}\mbox{\epsffile{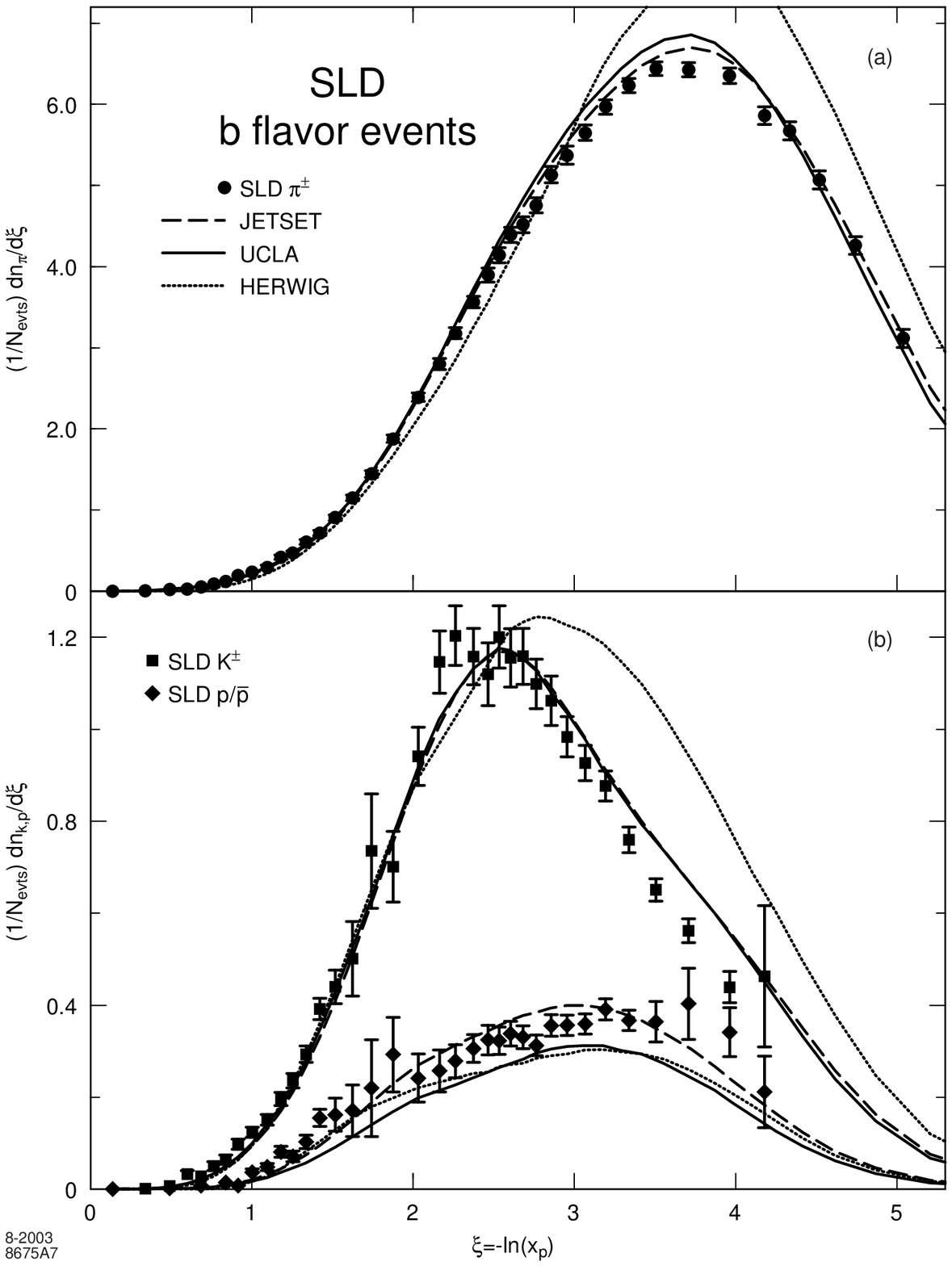}}\end{center}
 \vspace{-0.8cm}
 \caption{ \baselineskip=12pt  \label{xsbbmc}
Distributions of $\xi$ in $b$-flavor events for (a) pions (circles) and
(b) kaons (squares) and protons (diamonds), compared with the predictions of
the three models.
    }
\end{figure}

\begin{figure}
 \vspace{-0.2cm}
  \epsfxsize=6.5in
  \begin{center}\mbox{\epsffile{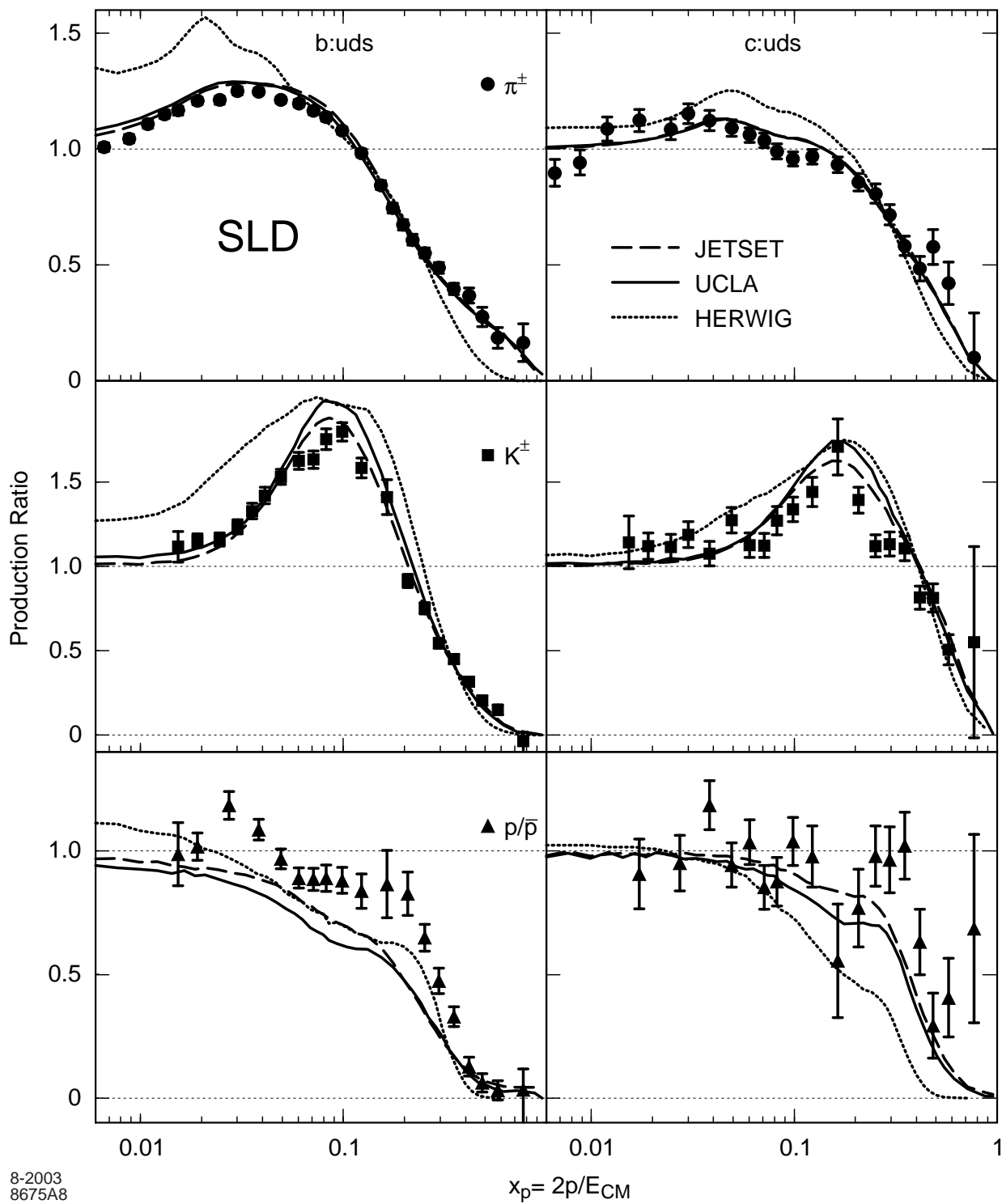}}\end{center}
 \vspace{-0.8cm}
 \caption{ \baselineskip=12pt  \label{xsrbumc}
Ratios of pion (top), kaon (middle) and proton (bottom) production in
$b$-flavor events to that in light-flavor events (left) and in
$c$-flavor:light-flavor events (right).
Some bins have been combined for clarity.
Also shown are the predictions of the three models.
    }
\end{figure}

In figs.~\ref{xsudsmc}--\ref{xsbbmc} we show the $\xi$ distributions for the
three flavor categories, and in fig.~\ref{xsrbumc} we show the ratios
of production in $b$-flavor to light-flavor events and $c$-flavor to
light-flavor events vs. $x_p$.
At low momentum (high $\xi$),
there is substantially higher production of charged pions in
$b$- and $c$-flavor events than in light-flavor events,
and much higher production of charged kaons in $b$-flavor events than in
light- or $c$-flavor events;
proton production is roughly equal at low momentum.
As momentum increases, the production of all three charged hadron species
falls much
more rapidly in $b$-flavor events than in light-flavor events, and that in
$c$-flavor events also drops off sharply at very low $\xi$.

\begin{table}
\begin{center}
 \begin{tabular}{|c|r@{$\pm$}r|r@{$\pm$}r|
                    r@{$\pm$}r|r@{$\pm$}l|r@{$\pm$}l|} \hline
& \multicolumn{6}{|c|}{ } & \multicolumn{4}{|c|}{ } \\ [-.3cm]
& \multicolumn{6}{|c|}{Charged Particle Production Rates} &
  \multicolumn{4}{|c|}{Ratios} \\
$<\!\!x_p\!\!>$
& \multicolumn{2}{|c|}{$u\bar{u}$, $d\bar{d}$, $s\bar{s}$}
& \multicolumn{2}{|c|}{$c\bar{c}$}
& \multicolumn{2}{|c|}{$b\bar{b}$}
& \multicolumn{2}{|c|}{$c$:$uds$}
& \multicolumn{2}{|c|}{$b$:$uds$} \\[.1cm]\hline
0.0066 & 507.8& 11.9&468.7& 25.3&546.1& 14.2 & 0.923&0.053 & 1.075&0.014 \\
0.0088 & 505.2&  9.2&485.4& 22.6&558.5& 10.3 & 0.961&0.050 & 1.105&0.013 \\
0.0109 & 465.3&  7.5&507.2& 20.2&531.9&  8.0 & 1.090&0.052 & 1.143&0.014 \\
0.0131 & 421.9&  6.3&464.0& 17.6&490.8&  6.3 & 1.100&0.050 & 1.163&0.014 \\
0.0153 & 371.7&  5.8&422.9& 16.9&436.5&  6.7 & 1.137&0.054 & 1.174&0.020 \\
0.0191 & 315.5&  4.2&349.1& 12.7&382.8&  4.4 & 1.107&0.049 & 1.213&0.017 \\
0.0246 & 250.5&  3.2&274.1&  9.5&308.6&  3.4 & 1.094&0.046 & 1.232&0.018 \\
0.0301 & 200.3&  2.3&231.5&  7.3&254.8&  2.6 & 1.156&0.044 & 1.272&0.017 \\
0.0356 & 167.3&  2.0&187.5&  6.9&213.2&  2.1 & 1.121&0.050 & 1.275&0.016 \\
0.0411 & 140.4&  1.6&162.3&  5.2&182.1&  1.9 & 1.156&0.044 & 1.297&0.017 \\
0.0465 & 121.2&  1.3&136.6&  4.2&154.5&  1.7 & 1.127&0.041 & 1.275&0.018 \\
0.0521 & 105.5&  1.1&117.3&  3.6&134.3&  1.6 & 1.112&0.040 & 1.273&0.019 \\
0.0576 &  91.2&  1.0& 99.1&  3.0&118.6&  1.4 & 1.087&0.038 & 1.301&0.020 \\
0.0630 & 81.29& 0.89&89.21& 2.65&102.4&  1.3 & 1.097&0.037 & 1.260&0.020 \\
0.0685 & 72.69& 0.81&78.25& 2.42&91.92& 1.22 & 1.077&0.037 & 1.265&0.021 \\
0.0740 & 65.92& 0.76&69.26& 2.23&83.63& 1.14 & 1.051&0.038 & 1.269&0.022 \\
0.0795 & 58.06& 0.70&62.25& 2.06&75.06& 1.10 & 1.072&0.040 & 1.293&0.024 \\
0.0850 & 53.26& 0.66&55.28& 1.93&66.58& 1.04 & 1.038&0.040 & 1.250&0.024 \\
0.0931 & 45.37& 0.48&49.14& 1.42&57.31& 0.74 & 1.083&0.035 & 1.263&0.020 \\
0.1040 & 38.55& 0.43&40.11& 1.29&47.80& 0.71 & 1.040&0.037 & 1.240&0.022 \\
0.1150 & 32.84& 0.41&35.64& 1.22&39.19& 0.67 & 1.085&0.041 & 1.194&0.025 \\
0.1310 & 26.05& 0.28&28.94& 0.83&29.54& 0.48 & 1.111&0.035 & 1.134&0.022 \\
0.1530 & 19.79& 0.29&21.99& 0.86&20.69& 0.48 & 1.111&0.048 & 1.045&0.029 \\
0.1750 & 15.75& 0.35&16.51& 1.03&15.36& 0.55 & 1.048&0.071 & 0.975&0.041 \\
0.1970 & 12.16& 0.17&12.69& 0.50&10.65& 0.26 & 1.044&0.046 & 0.876&0.024 \\
0.2189 & 10.27& 0.14&10.41& 0.40& 8.06& 0.20 & 1.013&0.044 & 0.785&0.022 \\
0.2410 &  8.14& 0.11& 7.86& 0.32& 6.28& 0.16 & 0.965&0.044 & 0.772&0.022 \\
0.2629 &  6.62& 0.10& 6.37& 0.28& 4.69& 0.13 & 0.961&0.047 & 0.709&0.023 \\
0.2849 & 5.565&0.087&5.060&0.241&3.490&0.110 & 0.909&0.048 & 0.627&0.022 \\
0.3068 & 4.428&0.076&4.080&0.210&2.935&0.098 & 0.922&0.052 & 0.663&0.025 \\
0.3338 & 3.588&0.057&3.123&0.156&2.041&0.068 & 0.870&0.048 & 0.569&0.021 \\
0.3666 & 2.706&0.049&2.141&0.126&1.534&0.058 & 0.791&0.051 & 0.567&0.024 \\
0.3997 & 2.062&0.042&1.472&0.106&1.111&0.051 & 0.714&0.055 & 0.539&0.027 \\
0.4325 & 1.631&0.037&0.952&0.090&0.736&0.042 & 0.584&0.058 & 0.451&0.028 \\
0.4651 & 1.193&0.034&0.935&0.085&0.510&0.037 & 0.783&0.076 & 0.427&0.033 \\
0.5035 & 0.912&0.026&0.485&0.061&0.330&0.027 & 0.532&0.070 & 0.362&0.032 \\
0.5470 & 0.632&0.023&0.372&0.055&0.188&0.022 & 0.588&0.091 & 0.298&0.037 \\
0.6083 & 0.398&0.015&0.140&0.031&0.089&0.012 & 0.353&0.081 & 0.224&0.031 \\
0.7047 & 0.172&0.011&0.027&0.016&0.017&0.006 & 0.158&0.096 & 0.099&0.034 \\
0.8383 & 0.027&0.005&0.011&0.010&0.003&0.002 & 0.412&0.386 & 0.103&0.095 \\
  [.1cm] \hline
  \end{tabular}
\caption{\baselineskip=12pt  \label{xsfch}
Differential production rates (1/N$_{evts}$)d$n_{chg}$/d$x_p$ of stable charged
particles per $Z^0$ decay into light, $c$ and $b$ primary flavors.
}
\end{center}
\end{table}

These features are consistent with expectations based on the known properties
of $e^+e^- \!\!\rightarrow \!\! b\bar{b}$ events, namely that a large fraction
of the event energy is carried by the leading $B$ and $\bar{B}$
hadrons~\cite{bfrag},
leaving little energy available to produce nonleading hadrons.
The $B$ hadrons decay into a large number of lighter particles, including
on average 5.5 stable charged hadrons \cite{pdg}, which are expected to
populate primarily the region $2<\xi<4$.
The mixture of particle types might be similar to that in
light-flavor jets, except that the weak decay chain of the $B$ hadron should
produce one `additional' $K^\pm$ per event, and baryon production might be
suppressed since a typical baryon-antibaryon pair mass is not small compared
with the $B$ hadron mass.
Similar effects in $c$-jets result in an intermediate situation:
each jet contains a charmed hadron with on average about half the beam
energy, a lower fraction than for $B$-hadrons,
which leaves more energy available for nonleading hadrons than in $b$-jets.
A $D$ meson decay produces an additional kaon that often carries a large
fraction of its momentum, and fewer additional particles than a $B$ decay.
Our results are consistent with those published previously~\cite{chgxsd} and
considerably more precise.

Also shown in figs.~\ref{xsudsmc}--\ref{xsrbumc} are the predictions of the
three models;
they all reproduce the observed flavor dependence qualitatively.
In the case of the light flavors, problems with the models very similar to those
seen in the preceding section for the flavor inclusive sample are observed,
indicating problems in the modelling of hadronization, and not just in that of
heavy hadron decay.
Some of the discrepancies are larger in the light flavors than in all flavors;
in particular the structure at small $\xi$ in the HERWIG model is now very
pronounced for all three particle types.

In $c$-flavor events, all models predict the pion yield within a few
percent, but the spectra are slightly too hard.
JETSET again predicts the shape of the proton spectrum reasonably well, but is
slightly high on the amplitude;
UCLA and HERWIG describe the data at large $\xi$, but fall well below the data
at small $\xi$.
JETSET and UCLA predict a kaon spectrum in which those from leading $D$ hadron
decays peak at $\xi\approx 2$, and those from the hadronization of the remaining
jet form a broad shoulder at larger $\xi$ values.
They are qualitatively consistent with the data, though the data prefer a lower
shoulder.
HERWIG predicts a quite different shape that is inconsistent with the data,
probably reflecting known problems in the modelling of heavy hadron production
and decay.

In $b$-flavor events, JETSET and UCLA describe the pion and kaon spectra
reasonably well, predicting a few too many pions in the range $2.5<\xi<4$, and
too many kaons for $\xi>3.5$;
both also predict the shape of the proton spectrum well, but UCLA is too low
in amplitude and JETSET is consistent with the data.
Again, HERWIG predicts very different spectra, all of which are inconsistent
with the data at large $\xi$ values.

The rate for all charged tracks in each of the flavor samples was
derived by summing the rates for the three charged hadron species and adding the
simulated contribution from leptons.
These rates and their ratios are listed in table~\ref{xsfch};
no plots are shown since the features correspond to those of the pion
data and models in figs.~\ref{xsudsmc}--\ref{xsrbumc}.
Our data are consistent with those published by DELPHI~\cite{chgxsd} and
OPAL~\cite{chgxso}, and substantially more precise.

\section{Comparison with QCD Predictions}

\begin{figure}
 \vspace{-1.cm}
  \epsfxsize=6.5in
  \begin{center}\mbox{\epsffile{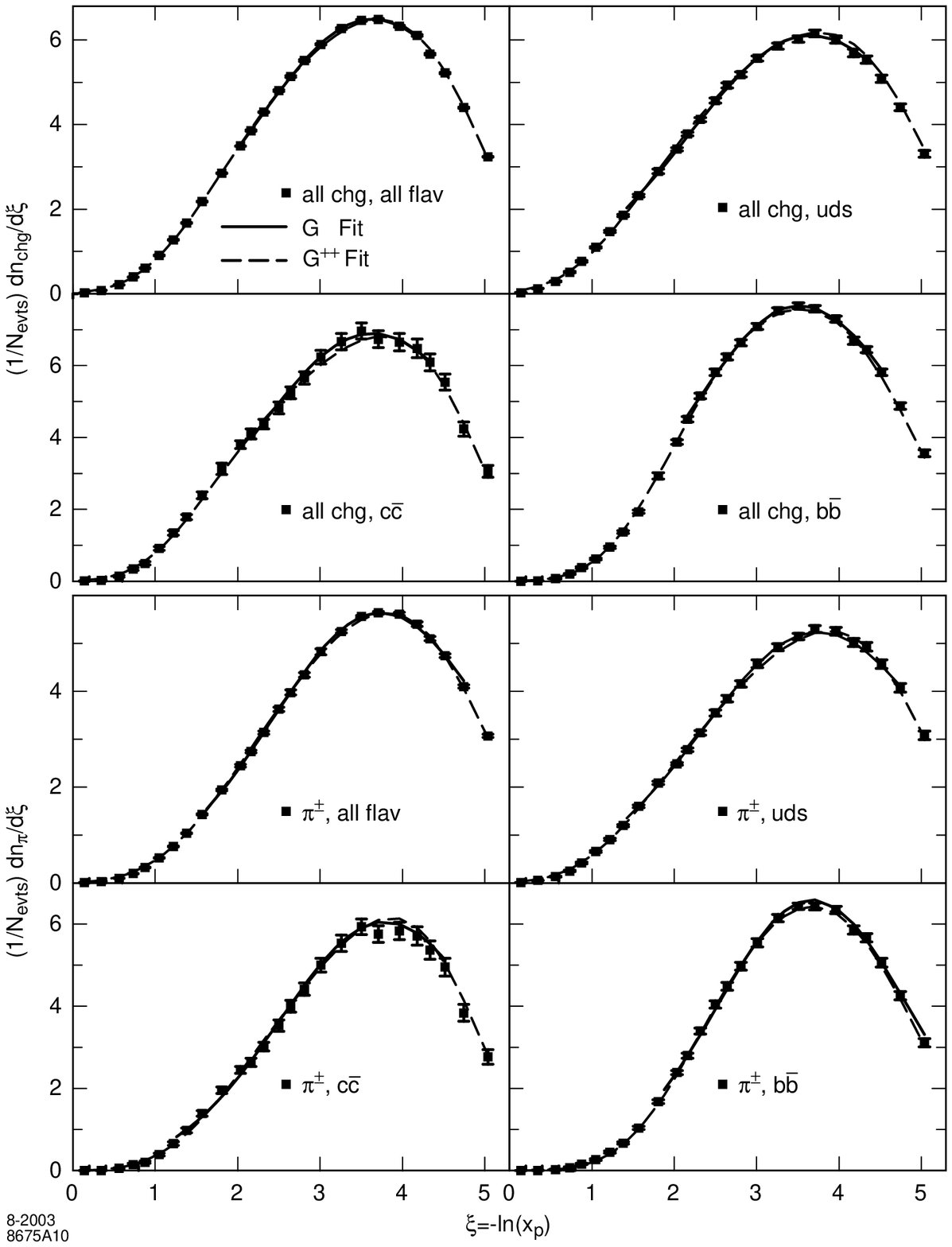}}\end{center}
 \vspace{-0.8cm}
 \caption{ \baselineskip=12pt  \label{xichpi}
Measured $\xi$ distributions for all charged particles (top) and pions (bottom)
in all-, light-, $c$- and $b$-flavor events.
Some bins have been combined for clarity.
The solid (dashed) lines represent the results of the maximal Gaussian ($G$)
and $G^{++}$ fits described in the text.
    }
\end{figure}

\begin{figure}
 \vspace{-1.cm}
  \epsfxsize=6.5in
  \begin{center}\mbox{\epsffile{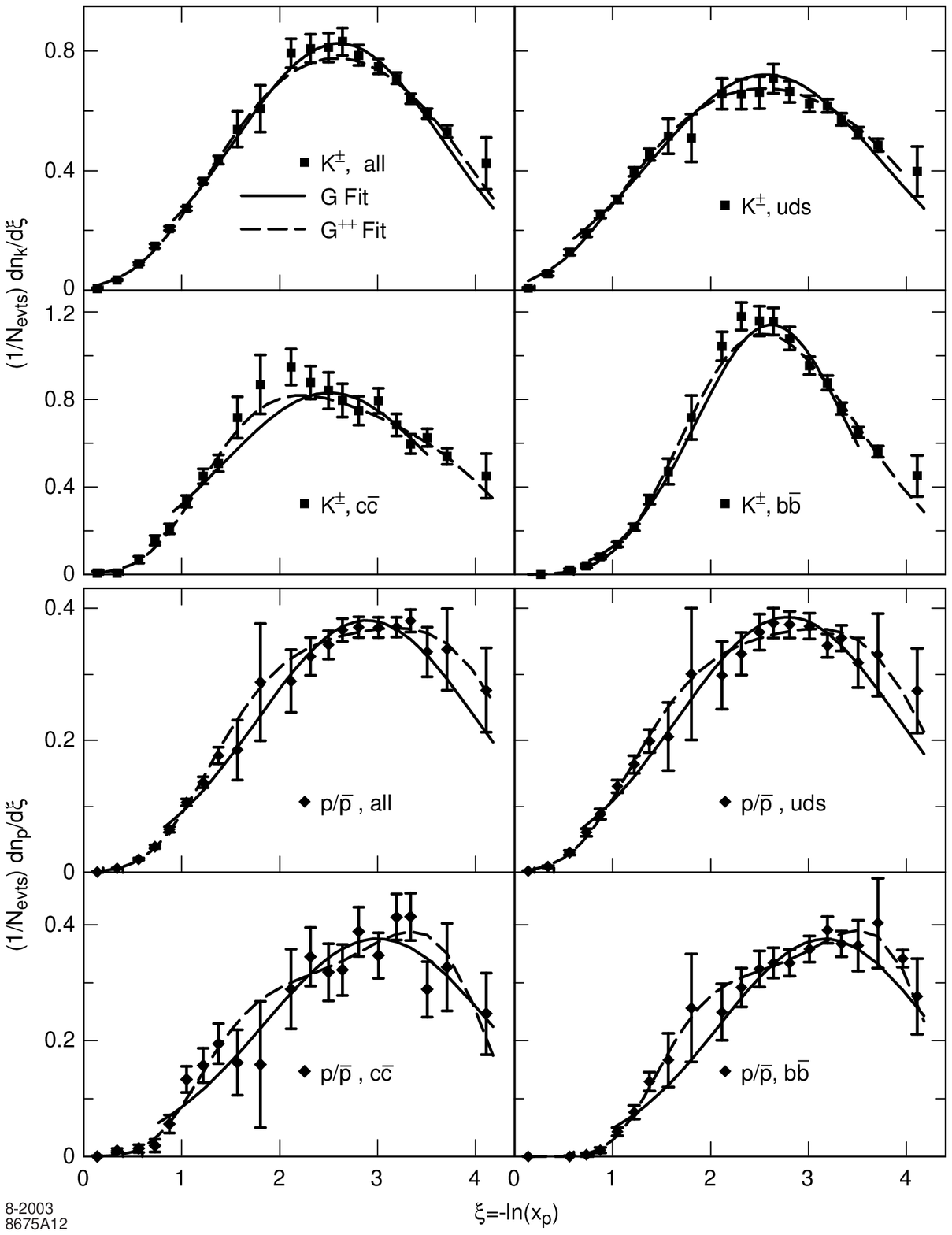}}\end{center}
 \vspace{-0.8cm}
 \caption{ \baselineskip=12pt  \label{xikapr}
Measured $\xi$ distributions for charged kaons (top) and protons (bottom)
in all-, light-, $c$- and $b$-flavor events.
Some bins have been combined for clarity.
The solid (dashed) lines represent the results of the maximal Gaussian ($G$)
and $G^{++}$ fits described in the text.
    }
\end{figure}

We tested the predictions of QCD in the Modified Leading Logarithm
Approximation (MLLA),
combined with the ansatz of Local Parton-Hadron Duality (LPHD),
by fitting Gaussian and distorted Gaussian functions to our measured $\xi$
distributions.
Examples of such fits are shown in figs.~\ref{xichpi} and \ref{xikapr}.
In each case, we first fitted a simple Gaussian to each set of points within a
region of $\xi$ at least 0.5 units in size centered on the approximate peak
position.
We chose the largest such symmetric range for which the confidence level of the
$\chi^2$ of the fit exceeded 0.01, and then added points on one side (in all
cases at lower $\xi$ values) as long as the confidence level remained above
0.01.
We found that the Gaussian function could describe the data over a symmetric
range of at least 0.8--1.6 units about the peak position, consistent with one
prediction of MLLA QCD; the fit could be extended toward lower $\xi$ by as much
as 1.7 additional units, although it must be noted that our errors on the kaon
and proton spectra are rather large in this range.
The largest fittable ranges are given in table~\ref{xistar} and the
corresponding fitted functions are shown on figs.~\ref{xichpi} and \ref{xikapr}.

We next introduced a skewness term into the function (G$^{+}$) and repeated the
above procedure.
We found that the symmetric range could be extended in only some cases, and by
at most 0.6 units (see table~\ref{xistar}).
In all-flavor and light-flavor events there was always an increase in the
fittable range on either the low or high $\xi$ end, and the fitted skewness
values were small.
In $c$- and $b$-flavor events, however, there were some cases in which no
increase in the fittable range could be obtained, and others in which the range
could be increased but the skewness value increased rapidly.

The addition of a kurtosis term (G$^{++}$) had similar results
(see table~\ref{xistar});
the fittable range was increased greatly, in many cases to the entire measured
range, however both the skewness and kurtosis values became large
for $c$- and $b$-flavor events.
The resulting $G^{++}$ functions are shown on figs.~\ref{xichpi}--\ref{xikapr}
over the entire $\xi$ range; the large distortions are evident in the
$c$- and $b$-flavor events for kaons and protons.
The MLLA prediction that a Gaussian with small distortion terms
should describe the data over a range substantially larger than one unit about
the peak position holds for the light flavors,
but does not hold for heavy flavors;
this might be expected since the calculation assumes massless partons.

The peak $\xi^*$ of the $\xi$ distribution is predicted to decrease
exponentially with increasing particle mass.
Following convention, we took the mean of the fitted Gaussian over a range of
one unit about the peak as an estimate of $\xi^*$, and,
in addition to statistical and experimental systematic errors,
we considered a variation of the fit range.
A fit was performed to each set of contiguous points with a smallest (largest)
$\xi$ value between 0.75 and 1.25 units below (above) the peak position.
Half the difference between the highest and lowest of the fitted means was taken
as an estimate of the systematic uncertainty due to the fit range.
The resulting values of $\xi^*$ are listed in table~\ref{xistar}, where a
considerable flavor dependence is seen.

\begin{table}
\begin{center}
\begin{tabular}{|r|c|c|c||r@{$\pm$}l@{$\pm$}l@{$\pm$}l|}\hline
  & \multicolumn{3}{c||}{  } & \multicolumn{4}{c|}{Peak position} \\[-.3cm]
  & \multicolumn{3}{c||}{Max. fit range} & \multicolumn{4}{c|}{  } \\
  & $G$ & $G^+$ & $G^{++}$ & $\xi$* & stat. & syst. & fit \\[.1cm] \hline
                  &          &          &          & \multicolumn{4}{c|}{  }\\[-.3cm]
      All         &1.62--4.74&1.88--5.21&0.92--5.21& 3.752&0.004&0.008&0.010\\[.1cm] 
$\pi^{\pm}$: $uds$&1.40--4.85&1.55--5.21&0.64--5.21& 3.787&0.008&0.027&0.014\\[.1cm] 
              $c$ &1.30--4.60&1.47--5.21&0.87--5.21& 3.726&0.014&0.090&0.039\\[.1cm]
              $b$ &1.81--5.21&1.81--5.21&0.00--5.21& 3.684&0.008&0.005&0.001
\\[.1cm] \hline
      All         &0.92--4.18&0.49--4.18&0.34--4.18& 2.549&0.018&0.042&0.062\\[.1cm] 
  $K^{\pm}$: $uds$&0.55--4.25&0.24--4.25&0.24--4.25& 2.592&0.032&0.091&0.021\\[.1cm] 
              $c$ &0.87--3.61&0.00--4.25&0.00--4.25& 2.412&0.030&0.018&0.012\\[.1cm]
              $b$ &0.72--3.61&0.00--4.25&0.00--4.25& 2.587&0.014&0.017&0.004
\\[.1cm] \hline
      All         &0.84--4.18&0.76--4.18&0.14--4.18& 3.084&0.101&0.056&0.031\\[.1cm] 
p$\bar{\rm p}$: $uds$&0.64--4.25&0.42--4.25&0.00--4.25& 2.859&0.076&0.028&0.026\\[.1cm] 
              $c$ &0.72--4.25&0.72--4.25&0.00--4.25& 3.079&0.113&0.023&0.068\\[.1cm]
              $b$ &0.97--4.25&0.97--4.25&0.00--4.25& 3.513&0.100&0.254&0.011
\\[.1cm] \hline
 \end{tabular}
\caption{ \baselineskip=12pt  \label{xistar}
The widest ranges in $\xi$ (see text) over which a Gaussian function ($G$) was
able to describe the data alone, and with the addition of skewness ($G^+$)
and kurtosis terms ($G^{++}$).
Peak positions $\xi^*$ from the Gaussian fits described in the text;
the errors are statistical, experimental systematic, and due to variation of
the fit range.}
\end{center}
\end{table}

\begin{figure}
 \vspace{-1.cm}
  \epsfxsize=6.5in
  \begin{center}\mbox{\epsffile{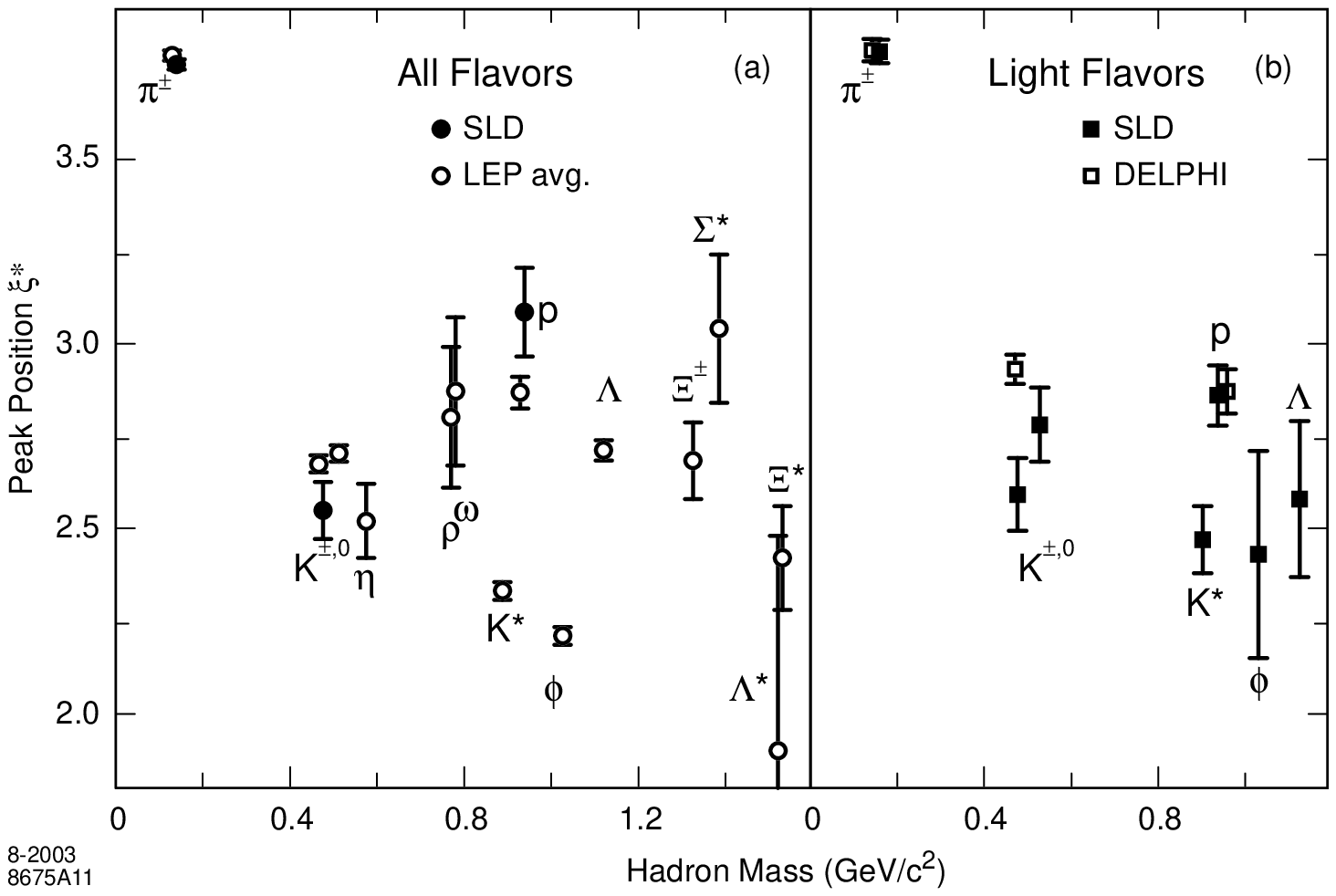}}\end{center}
  \vspace{-0.8cm}
 \caption{ \baselineskip=12pt  \label{xipeak}
Peak positions $\xi^*$ as a function of hadron mass in (a) all- and (b)
light-flavor events, along with previous results from experiments at the $Z^0$.
    }
\end{figure} 

The values in all-flavor and light-flavor events are shown in fig.~\ref{xipeak}
along with previous results~\cite{bohrer,bfp,chgxsa,chgxsd}.
As observed previously, the $\xi^*$ values for the measured hadron species in
all-flavor events do not fall on a single trajectory as a function of mass,
although roughly parallel
trajectories can be postulated for mesons and baryons.
For light flavors, the $\xi^*$ value for pions is quite similar to that for all
flavors, and those for the other mesons (baryons) tend to be higher (lower), as
would be expected if the MLLA prediction is correct for primary hadrons,
but the observed distributions are distorted by secondary particles.
However it is still not possible to draw a single trajectory through all the
light-flavor points and this apparent deficiency in the the theory remains;
it is possible that
the exclusion of additional non-primary hadrons might remove it.

Our previous results~\cite{bfp} were used in conjunction with data from
other experiments to test the perturbative QCD calculations of the $E_{CM}$
dependence of the rates at high-$x_p$~\cite{kkp}.
The results presented here can be used to make a slightly more precise test;
however more precise data at lower and/or higher $E_{CM}$ are needed to confront
these calculations in a more meaningful way.


\section{Total Production Rates}

We have integrated our differential production rates over their respective
measurement ranges, taking into account the bin-to-bin correlations in the
systematic errors.  These integrated rates per event are listed in
tables \ref{chgxs}--\ref{pfraxa} and \ref{xsfpi}--\ref{xsfpr};
the errors are dominated by overall normalization uncertainties corresponding
to the uncertainty in our track reconstruction efficiency.
In order to quote total rates, we must extrapolate into the unmeasured
regions, $\xi>5.21$ for inclusive charged particles and pions,
and $\xi>4.25$ for kaons and protons.
As can be seen in figs.~\ref{xsallmc}--\ref{xsbbmc}, and
\ref{xichpi}--\ref{xikapr}, this is a non-trivial effect, and none of the models
or functions provides an ideal estimate of the accepted fraction.
However, a set of four estimates derived from the three models and the $G^{++}$
function might be considered to cover a reasonable range of possibilities.
For the $c$- and $b$-flavor events, we used in addition the generator used for
our detector simulation~\cite{sldsim}, a version of JETSET tuned to the world's
data on $D$ and $B$ hadron production and decay.

\begin{table}
\begin{center}
\begin{tabular}{|r|r@{$\pm$}l|r@{$\pm$}l||l|r@{$\pm$}l|}\hline
  & \multicolumn{2}{c|}{  }         & \multicolumn{2}{c||}{  }
  & \multicolumn{3}{c|}{  }           \\[-.4cm]
  & \multicolumn{2}{c|}{Coverage}   & \multicolumn{2}{c||}{Yield/Event}
  & \multicolumn{3}{c|}{Difference}   \\[.05cm]\hline
  & \multicolumn{2}{c|}{  }         & \multicolumn{2}{c||}{  }
 && \multicolumn{2}{c|}{  }           \\[-.4cm]
      All & 0.925 & 0.007 & 17.007 & 0.209&&\multicolumn{2}{c|}{ } \\[0.05cm]
$\pi^\pm$:
    $uds$ & 0.923 & 0.008 & 16.579 & 0.304 & $c\!-\!uds$ &   0.375 & 0.587 \\[0.05cm]
      $c$ & 0.931 & 0.013 & 16.954 & 0.556 & $b\!-\!uds$ &   1.557 & 0.232 \\[0.05cm]
      $b$ & 0.929 & 0.007 & 18.136 & 0.330 & $b\!-\!c$   &   1.182 & 0.498
\\[.1cm] \hline
  & \multicolumn{2}{c|}{  }         & \multicolumn{2}{c||}{  }
 && \multicolumn{2}{c|}{  }           \\[-.4cm]
      All & 0.941 & 0.010 &  2.203 & 0.071&&\multicolumn{2}{c|}{ } \\[0.05cm]
 $K^\pm$:
   $uds$  & 0.934 & 0.007 &  2.000 & 0.068 & $c\!-\!uds$ &   0.427 & 0.074 \\[0.05cm]
     $c$  & 0.937 & 0.005 &  2.427 & 0.100 & $b\!-\!uds$ &   0.510 & 0.037 \\[0.05cm]
     $b$  & 0.947 & 0.006 &  2.510 & 0.086 & $b\!-\!c$   &   0.083 & 0.075
\\[.1cm] \hline
  & \multicolumn{2}{c|}{  }         & \multicolumn{2}{c||}{  }
 && \multicolumn{2}{c|}{  }           \\[-.4cm]
      All & 0.933 & 0.003 &  1.054 & 0.035&&\multicolumn{2}{c|}{ } \\[0.05cm]
p$\bar{\rm p}$:
    $uds$ & 0.921 & 0.010 &  1.094 & 0.043 & $c\!-\!uds$ &$-$0.060 & 0.074 \\[0.05cm]
      $c$ & 0.899 & 0.038 &  1.034 & 0.077 & $b\!-\!uds$ &$-$0.091 & 0.034 \\[0.05cm]
      $b$ & 0.906 & 0.020 &  1.004 & 0.046 & $b\!-\!c$   &$-$0.031 & 0.074
\\[.1cm] \hline 
  & \multicolumn{2}{c|}{  }         & \multicolumn{2}{c||}{  }
 && \multicolumn{2}{c|}{  }           \\[-.4cm]
    $uds$ & 0.927 & 0.007 & 20.048 & 0.316 & $c\!-\!uds$ &   1.048 & 0.718 \\[0.05cm]
All Chg.:
     $c$  & 0.940 & 0.015 & 21.096 & 0.653 & $b\!-\!uds$ &   3.050 & 0.311 \\[0.05cm]
     $b$  & 0.933 & 0.010 & 23.098 & 0.378 & $b\!-\!c$   &   2.002 & 0.643
\\[.1cm] \hline 
 \end{tabular}
\caption{ \baselineskip=12pt  \label{xsint}
Estimated fractions of the total production rates contained within the range
of the measurements;
corrected yields of charged hadrons per $Z^0$ decay into events of each flavor
category.
The rightmost columns show differences between flavors, for which some
uncertainties cancel.}
\end{center}
\end{table}

From the pions and protons generated using each of the models, we
calculated the fraction that were generated in the range of our
measurement, as the predictions, when normalized to the data, provided
reasonable descriptions of the shape.
For the charged kaons (figs.~\ref{xsallmc}--\ref{xsbbmc}), all models predict
spectra that are too soft;  we therefore scaled the predictions along the $\xi$
axis until the best agreement with the data was found.
This procedure changed the acceptance correction by a few percent for $b$- and
$c$-flavor events, and by as much as 12\% for the HERWIG model in light-flavor
events.
The fitted $G^{++}$ function was integrated to obtain another estimate of the
accepted fraction in those cases where it gave an acceptable $\chi^2$ over the
entire measured range.
Otherwise, it was used to calculate a fraction above the lower bound given in
table~\ref{xistar}, and the data below that bound were added to obtain an
overall fraction.

For each hadron species and flavor category these four or five estimates of the
fraction were found to be similar, with a typical rms value of about 1\%
relative.
We took their average as our central value of the fraction in each case, and
took their rms as the uncertainty due to the extrapolation procedure.
These values along with the corrected total rates are listed in
table~\ref{xsint};  also shown are differences between the three flavor
categories, for which some of the uncertainties cancel.
The results were found to be consistent with previous measurements; the
precision is similar for all-flavor events, and substantially improved for
light- and $b$-flavors; ours are the only measurements for $c$-flavor events.

Roughly 10\% more pions are produced in $b$-flavor events than in $c$- or
light-flavor events;
roughly 20\% more kaons are produced in both $b$- and $c$-flavor
events than in light-flavor events;
roughly 10\% fewer protons are produced in $b$- than in light-flavor events.
The total charged multiplicities and differences between flavors
are consistent with previous, dedicated measurements~\cite{chmult,chgxso};
they have comparable precision with different systematic error sources.

\section{Leading Particle Effects}

We extended these studies to look for differences between particle and
antiparticle production in light quark (as opposed to antiquark) jets,
in order to address the question of whether e.g. 
a primary $u$-initiated jet contains more hadrons that contain a valence
$u$-quark
(e.g. $\pi^+$, $K^+$, p) than hadrons that do not
(e.g. $\pi^-$, $K^-$, $\bar{\rm p}$).
To this end we used the light
quark- and antiquark-tagged hemispheres described in section 3.

We measured the differential production rates per light quark jet
\begin{eqnarray}
R^{q}_{h} &=& {1\over{2N_{evts}}}{d\over{dx_{p}}}\left[ N(q\rightarrow
h)+N(\bar{q}\rightarrow\bar{h})\right],\\
R^{q}_{\bar{h}} &=& {1\over{2N_{evts}}}{d\over{dx_{p}}}\left[
N(q\rightarrow\bar{h})+N(\bar{q}\rightarrow h)\right],
\end{eqnarray}
where: $q$ and $\bar{q}$ represent light-flavor quark and antiquark jets
respectively; $N_{evts}$ is the total number of events in the sample; $h$
represents $\pi^{+}$, $K^{+}$ or p, and $\bar{h}$ the corresponding antihadron.
Then, for example, $N(q\rightarrow h)$ is
the number of hadrons of species $h$ in light quark jets.
This formulation assumes CP symmetry, i.e.
$N(q\rightarrow h) = N(\bar{q} \rightarrow \bar{h})$, which was found to be
satisfied in the data in all cases.

The charged hadron analysis was repeated on the sample of positively
charged tracks in the quark-tagged jets and negatively charged tracks in the
antiquark-tagged jets, yielding measured values of
$R^{q}_{\pi^{+}}$, $R^{q}_{K^{+}}$, and $R^{q}_{\rm p}$ in the tagged samples.
The same procedure applied to the remaining tracks yielded $R^{q}_{\pi^{-}}$,
$R^{q}_{K^{-}}$, and $R^{q}_{\bar{\rm p}}$.

The decays of the leading heavy hadrons in simulated heavy flavor background
events give rise to differences between hadron and antihadron production in
the quark-tagged sample over the entire $x_p$ range, including an effect in
high-momentum pions of opposite sign to that seen in the light-flavor data
(below).
It is essential to understand and/or suppress this contribution;  our
simulation has been tuned to the available data on charmed and bottom hadron
decays, and our light-flavor tag reduces this background to
9\% $c\bar{c}$ and 2\% $b\bar{b}$ events.
At this level the simulated contribution to any difference is well below the
other uncertainties.
The simulated contribution to each rate was applied as a correction,
yielding differential production rates per light-quark-tagged jet.
For each hadron species, differential production rates in light quark
jets were then extracted by correcting for the light-tag bias (see sec. 6) and
unfolding for the effective quark (vs. antiquark) purity.
This purity depends on polar angle, and hence on the acceptance of the CRID
as well as the event selection discussed in section 3;
its average value was estimated from the simulation to be 0.72 for the
selected track sample.

\begin{table}
\begin{center}
 \begin{tabular}{|c|r@{$\pm$}l|r@{$\pm$}l|r@{$\pm$}r||
                  c|r@{$\pm$}l|r@{$\pm$}l|r@{$\pm$}r|}\hline
&\multicolumn{2}{|c|}{ }&\multicolumn{2}{|c|}{ }&\multicolumn{2}{|c||}{ }&
&\multicolumn{2}{|c|}{ }&\multicolumn{2}{|c|}{ }&\multicolumn{2}{|c|}{ }\\[-.3cm]
 $<\!\! x_p\!\! >$
&\multicolumn{2}{|c|} {$R_{\pi^+}$}&\multicolumn{2}{|c|}{$R_{\pi^-}$}
&\multicolumn{2}{|c||}{$D_{\pi^-}$(\%)}  &  $<\!\! x_p\!\! >$
&\multicolumn{2}{|c|} {$R_{\pi^+}$}&\multicolumn{2}{|c|}{$R_{\pi^-}$}
&\multicolumn{2}{|c|} {$D_{\pi^-}$(\%)} \\[.1cm]\hline
0.0066&243.9&4.0 &237.5&4.0 &--1.3&1.5&0.1150&12.25&0.30 &11.99&0.30 &--1.1&2.2\\
0.0088&236.7&3.6 &235.5&3.5 &--0.3&1.3&0.1311&9.454&0.188&9.663&0.189&  1.1&1.8\\
0.0110&210.5&3.1 &215.3&3.2 &  1.1&1.3&0.1530&7.231&0.163&7.386&0.164&  1.1&2.0\\
0.0131&194.5&2.9 &189.4&2.8 &--1.3&1.3&0.1750&5.594&0.141&5.433&0.139&--1.5&2.3\\
0.0153&168.1&2.6 &166.4&2.6 &--0.5&1.4&0.1970&4.330&0.123&4.360&0.122&  0.4&2.6\\
0.0191&137.8&1.5 &142.8&1.6 &  1.8&1.0&0.2189&3.302&0.108&3.503&0.108&  3.0&2.9\\
0.0246&108.1&1.4 &109.5&1.4 &  0.7&1.1&0.2410&2.619&0.092&2.721&0.093&  1.9&3.1\\
0.0301&84.99&1.18&88.51&1.19&  2.0&1.2&0.2629&2.059&0.082&2.142&0.083&  2.0&3.6\\
0.0356&68.97&1.06&71.82&1.07&  2.0&1.3&0.2850&1.623&0.074&1.844&0.075&  6.4&3.9\\
0.0411&57.26&0.96&59.36&0.97&  1.8&1.5&0.3070&1.294&0.065&1.378&0.065&  3.2&4.4\\
0.0466&49.27&0.91&49.63&0.91&  0.4&1.6&0.3337&0.974&0.048&1.161&0.049&  8.8&4.1\\
0.0521&43.81&0.86&42.11&0.86&--2.0&1.8&0.3665&0.721&0.042&0.908&0.044& 11.5&4.8\\
0.0576&37.47&0.79&36.45&0.78&--1.4&1.9&0.3996&0.523&0.036&0.637&0.037&  9.9&5.7\\
0.0631&32.00&0.72&32.79&0.72&  1.2&2.0&0.4326&0.424&0.032&0.451&0.032&  3.1&6.7\\
0.0685&28.67&0.68&28.02&0.67&--1.1&2.1&0.4653&0.278&0.028&0.375&0.030& 15.0&8.1\\
0.0740&24.88&0.63&26.16&0.63&  2.5&2.2&0.5034&0.221&0.021&0.222&0.022&  0.2&8.8\\
0.0795&21.33&0.59&24.09&0.60&  6.1&2.4&0.5464&0.107&0.018&0.192&0.020&  28.&11.\\
0.0850&20.51&0.56&20.15&0.56&--0.9&2.5&0.6085&0.066&0.011&0.112&0.012&  26.&12.\\
0.0931&17.82&0.37&17.17&0.36&--1.8&1.9&0.7046&0.020&0.007&0.054&0.009&  45.&20.\\
0.1041&14.52&0.33&14.61&0.33&  0.3&2.0&0.8342&0.011&0.004&0.006&0.004&--27.&40.\\
  [.1cm] \hline
  \end{tabular}
\caption{ \baselineskip=12pt  \label{xsqqpi}
Differential production rates $R_h^q=(1/2N_{evts}) dn_h/dx_p$ for 
positively and negatively charged pions $h=\pi^+$, $\pi^-$ 
in light ($u$, $d$ and $s$) quark jets from hadronic $Z^0$ decays, along with
the normalized difference
$D_{\pi^-}=(R_{\pi^-}^q-R_{\pi^+}^q)/(R_{\pi^-}^q+R_{\pi^+}^q)$.
The errors are the sum in quadrature of statistical errors and those systematic
errors arising from the light quark tagging and unfolding procedure.}
\end{center}
\end{table}

\begin{table}
\begin{center}
 \begin{tabular}{|c|r@{$\pm$}l|r@{$\pm$}l|r@{$\pm$}r||
                    r@{$\pm$}l|r@{$\pm$}l|r@{$\pm$}r|}\hline
&\multicolumn{2}{|c|}{ }&\multicolumn{2}{|c|}{ }&\multicolumn{2}{|c||}{ }
&\multicolumn{2}{|c|}{ }&\multicolumn{2}{|c|}{ }&\multicolumn{2}{|c|}{ }\\[-.3cm]
 $<\!\! x_p\!\! >$
&\multicolumn{2}{|c|} { $R_{K^+}$ }&\multicolumn{2}{|c|}{  $R_{K^-}$    }
&\multicolumn{2}{|c||}{ $D_{K^-}$ (\%)}
&\multicolumn{2}{|c|} {$R_{\rm p}$}&\multicolumn{2}{|c|}{$R_{\bar{\rm p}}$}
&\multicolumn{2}{|c|} {$D_{\rm p}$ (\%)} \\[.1cm]\hline
0.0153& 13.69&1.39 &12.89&1.39 &--3.0&9.4&  8.09&1.19 & 6.67&1.18 &  10.&14.\\
0.0191&  9.23&0.45 &10.10&0.45 &  4.5&4.1&  8.98&0.62 & 8.82&0.62 &  0.9&5.9\\
0.0246& 10.10&0.44 & 9.73&0.44 &--1.9&4.0&  7.42&0.72 & 6.41&0.72 &  7.4&9.3\\
0.0301&  8.67&0.40 & 8.87&0.40 &  1.1&4.1&  5.81&0.42 & 5.05&0.42 &  7.0&6.6\\
0.0356&  8.55&0.38 & 7.80&0.38 &--4.6&4.2&  5.52&0.34 & 5.19&0.34 &  3.1&5.6\\
0.0411&  7.91&0.38 & 8.08&0.38 &  1.1&4.3&  4.40&0.29 & 4.18&0.28 &  2.6&6.0\\
0.0466&  7.81&0.40 & 7.37&0.40 &  2.9&4.7&  3.74&0.26 & 3.40&0.26 &  4.8&6.6\\
0.0521&  5.27&0.40 & 7.04&0.41 & 14.3&5.9&  4.11&0.27 & 3.74&0.27 &  4.7&6.3\\
0.0576&  5.46&0.37 & 5.59&0.38 &  1.2&6.2&  3.48&0.25 & 3.32&0.25 &  2.4&6.6\\
0.0631&  4.91&0.35 & 5.69&0.35 &  7.4&6.0&  3.01&0.23 & 3.01&0.24 &--0.1&7.7\\
0.0685&  5.24&0.34 & 4.96&0.34 &--2.9&6.0&  2.87&0.23 & 2.47&0.23 &  7.6&7.7\\
0.0740&  4.65&0.33 & 5.14&0.34 &  5.1&6.2&  2.58&0.23 & 2.58&0.22 &  0.1&7.9\\
0.0795&  4.09&0.32 & 4.31&0.33 &  2.6&7.0&  2.34&0.22 & 1.94&0.21 &  9.2&9.1\\
0.0850&  4.31&0.32 & 3.84&0.33 &--5.9&7.3&  1.87&0.22 & 2.50&0.23 &--15.&10.\\
0.0931&  3.13&0.22 & 3.89&0.23 & 10.9&5.8&  1.94&0.16 & 1.87&0.16 &  2.1&7.5\\
0.1041&  3.07&0.22 & 3.28&0.23 &  3.3&6.4&  1.43&0.16 & 1.62&0.16 &--6.1&9.5\\
0.1150&  3.00&0.24 & 3.05&0.25 &  0.8&7.3&  1.45&0.18 & 1.33&0.17 &   4.&11.\\
0.1311&  2.02&0.18 & 2.91&0.19 & 17.9&6.8&  1.30&0.15 & 0.96&0.14 &  15.&12.\\
0.1530&  1.50&0.21 & 1.77&0.22 &   8.&12.&  0.88&0.19 & 1.07&0.20 &--10.&18.\\
0.1750&  1.09&0.28 & 1.77&0.30 &  23.&19.&  1.00&0.27 & 0.82&0.28 &  10.&28.\\
0.1970& 0.997&0.104&1.528&0.111& 21.0&6.9& 0.501&0.087&0.417&0.087&   9.&18.\\
0.2189& 0.856&0.078&1.593&0.086& 30.1&5.9& 0.612&0.064&0.374&0.063&  24.&13.\\
0.2410& 0.784&0.061&1.180&0.067& 20.1&5.9& 0.469&0.050&0.384&0.050&  10.&11.\\
0.2629& 0.571&0.054&1.105&0.060& 31.9&6.2& 0.475&0.045&0.275&0.043&  27.&11.\\
0.2850& 0.508&0.049&0.959&0.055& 30.7&6.4& 0.460&0.040&0.191&0.038&  41.&11.\\
0.3070& 0.366&0.043&0.824&0.049& 38.4&7.0& 0.321&0.033&0.174&0.032&  30.&12.\\
0.3337& 0.320&0.033&0.711&0.038& 27.9&6.2& 0.298&0.025&0.147&0.024&  34.&10.\\
0.3665& 0.249&0.029&0.529&0.033& 35.9&7.1& 0.198&0.020&0.123&0.020&  23.&11.\\
0.3996& 0.194&0.027&0.503&0.031& 44.4&7.4& 0.163&0.017&0.095&0.017&  26.&12.\\
0.4326& 0.167&0.024&0.372&0.027& 37.9&8.5& 0.114&0.015&0.083&0.014&  16.&13.\\
0.4653& 0.132&0.022&0.294&0.025& 38.1&9.9& 0.105&0.013&0.036&0.012&  49.&16.\\
0.5034& 0.089&0.018&0.267&0.021& 50.0&9.5& 0.085&0.010&0.022&0.009&  59.&15.\\
0.5464& 0.050&0.016&0.208&0.019&  61.&12.& 0.057&0.008&0.011&0.007&  67.&20.\\
0.6085& 0.063&0.011&0.118&0.012&  30.&11.& 0.024&0.004&0.009&0.004&  48.&21.\\
0.7046& 0.022&0.008&0.056&0.009&  43.&20.& 0.006&0.002&0.009&0.002&--18.&29.\\
0.8342& 0.005&0.005&0.006&0.005&   9.&83.& 0.000&0.001&0.003&0.001&--164&144\\
  [.1cm] \hline
  \end{tabular}
\caption{ \baselineskip=12pt  \label{xsqqkp}
Differential production rates $R_h^q=(1/2N_{evts}) dn_h/dx_p$ for 
positively and negatively charged kaons and protons
in light ($u$, $d$ and $s$) quark jets from hadronic $Z^0$ decays, along with
the normalized differences $D_{h}=(R_h^q-R_{\bar{h}}^q)/(R_h^q+R_{\bar{h}}^q)$.
The errors are the sum in quadrature of statistical errors and those systematic
errors arising from the light quark tagging and unfolding procedure.}
\end{center}
\end{table}

\begin{figure}
 \vspace{-0.2cm}
  \epsfxsize=6.5in
  \begin{center}\mbox{\epsffile{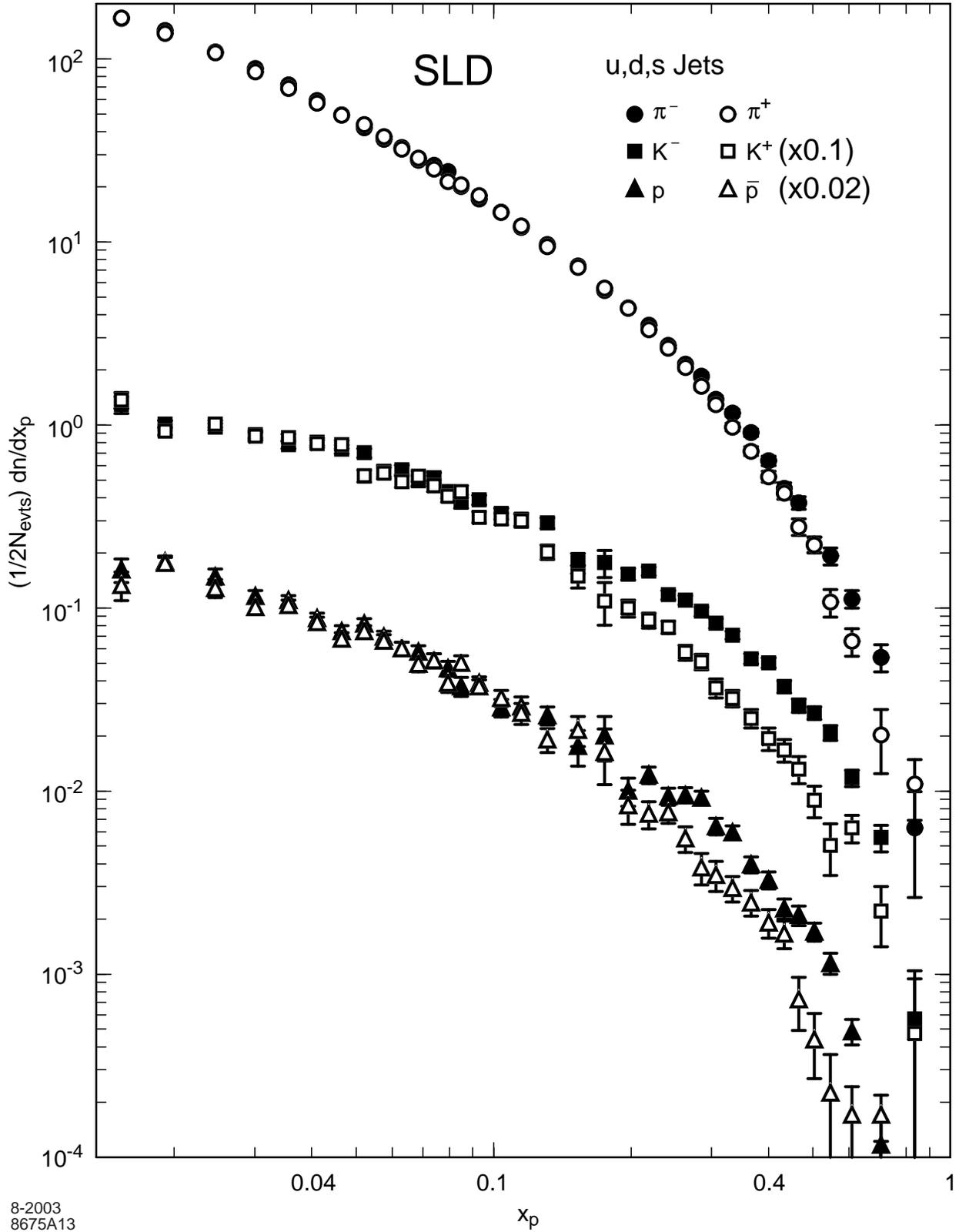}}\end{center}
 \vspace{-0.8cm}
 \caption{ \baselineskip=12pt  \label{xsqqa}
Fully corrected production rates for positively and negatively charged hadrons
in light quark jets.
The error bars include statistics and those systematic uncertainties that affect
the difference between the charges.
    }
\end{figure}

The measured differential production rates per light
quark jet are listed in
tables \ref{xsqqpi}--\ref{xsqqkp} and shown in fig.~\ref{xsqqa};
as for the flavor dependent results (sec. 6), the errors given are the sum in
quadrature of the statistical error and those systematic errors arising from
the tagging and correction procedures.
The latter include variation of:
the event tagging efficiencies and biases as described in section 6;
the electroweak parameters $R_b$, $R_c$, $A_b$ and $A_c$ by the
errors on their respective world average values;
the effective quark purity by $\pm$0.01;
the sum of $h$ and $\bar{h}$ rates in $c$- and $b$-flavor events by a
smooth parametrization of the errors in tables~\ref{xsfpi}--\ref{xsfpr};
and their difference by $\pm$20\% of itself
to cover the uncertainty in the electron beam polarization and statistical error
on the simulated purities.
The systematic errors are small compared with the statistical errors, and are
typically dominated by the uncertainty on the effective quark purity.

\begin{figure}
 \vspace{-1.cm}
  \epsfxsize=6.5in
  \begin{center}\mbox{\epsffile{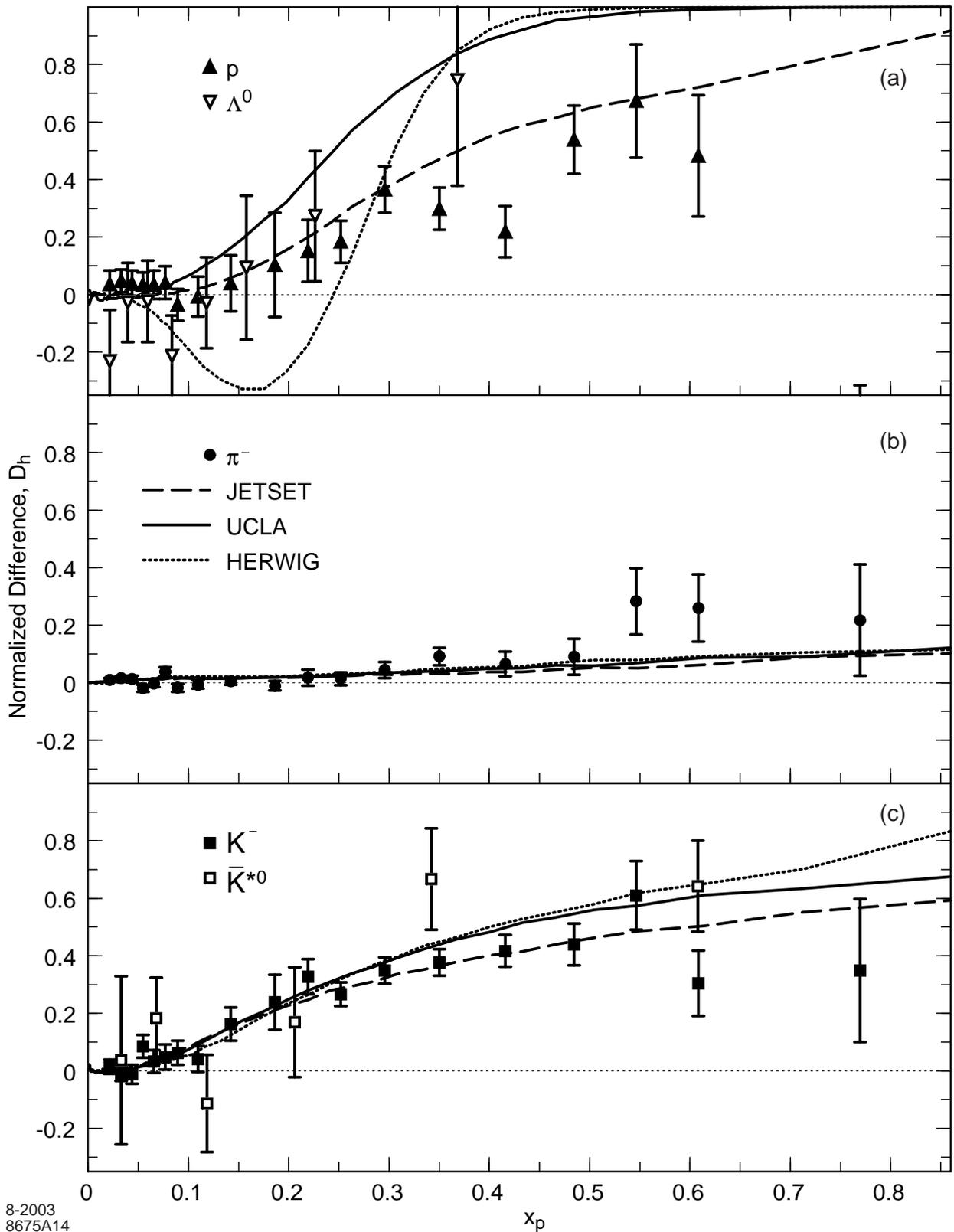}}\end{center}
 \vspace{-0.8cm}
 \caption{ \baselineskip=12pt  \label{ndmc}
Normalized differences between hadron and antihadron production in light quark
jets, compared with the predictions of the three models.
Some bins have been combined for clarity.
Our previous results for $\Lambda^0$ ($K^{*0}$) are overlaid on the proton
(kaon) plot; the corresponding model predictions are similar.
    }
\end{figure}

In all cases the hadron and antihadron rates are consistent at low
$x_p$.
For kaons (protons), significant differences appear for $x_p$ above about
0.15 (0.2) units and grow with increasing $x_p$.
For pions, differences are smaller, but significant for $x_p$ above about 0.3,
also appearing to grow with $x_p$.
It is convenient to show these results in the form of the difference between
hadron $h$ and antihadron $\bar{h}$ production normalized by the sum:
\begin{equation}
D_{h} =  {R^{q}_{h} - R^{q}_{\overline{h}}\over
          R^{q}_{h} + R^{q}_{\overline{h}}}.
\end{equation}
The common systematic errors cancel explicitly in this variable, which
is shown for the hadrons $h=\pi^-,K^-$, p in fig. \ref{ndmc} and listed in
tables \ref{xsqqpi}--\ref{xsqqkp}.
A value of zero corresponds to equal production of hadron and antihadron,
whereas a value of $+$(--)1 corresponds to complete dominance of
(anti)particle production in light quark jets.

The results for the protons afford the most straightforward interpretation.
Since baryons contain valence quarks and not antiquarks, the positive values
of $D_{\rm p}$ for $x_p>0.2$ are clear evidence for the production of leading
protons.
The data are consistent with a steady increase with $x_p$ to a plateau of
$D_{\rm p}\approx 0.5$ for $x_p>0.5$, although the errors on the highest $x_p$
points are quite large.
For $x_p<0.1$ the data are consistent, within common systematic errors, with
equal production of baryons and antibaryons, however it must be noted that the
contribution from non-leading hadrons is very high in this region and we
cannot exclude that some leading baryons are produced at low $x_p$.

The interpretation of our results for mesons is more complicated, since a meson
contains one valence quark along with one valence antiquark.
In SM $Z^0$ decays all primary down-type quarks are produced equally and with
the same forward-backward asymmetry, so that if a leading neutral
hadron such as $\bar{K}^{*0}$ ($s\bar{d}$) were produced equally in $s$ and
$\bar{d}$ jets then we would measure $D_{\bar{K}^{*0}}$ to be zero.
We previously measured~\cite{bfp} a significantly positive value (see
fig.~\ref{ndmc}) at high $x_p$,
indicating both that there is leading $\bar{K}^{*0}$ production
{\it and} that more leading $\bar{K}^{*0}$ are produced in $s$ jets than in
$\bar{d}$ jets.
This is an expected consequence of strangeness suppression in the
hadronization process.
That is, it is expected to be less likely for an $s\bar{s}$ pair to pop out of
the vacuum and the $s$ to pair up with an initial $\bar{d}$ than it is for 
a $d\bar{d}$ pair to pop out and the $\bar{d}$ to pair up with the initial $s$.

In the case of charged mesons such as $\pi^-$ ($d\bar{u}$), there is a
nonzero dilution of leading particle effects because of 
the different $Z^0$ branching ratios and forward-backward asymmetries of up- and
down-type quarks.
Assuming SM couplings to the $Z^0$, equal production of leading $\pi^+$
in $u$-jets and $\pi^-$ in $d$-jets, and no contribution from $s$-jets,
we calculate a dilution factor for this analysis of 0.27.
That is, we would expect to observe
$D^{obs}_{\pi^-} = 0.27 D^{true}_{d\rightarrow \pi^-}
                 = 0.27 D^{true}_{u\rightarrow \pi^+}$.

Our measured $D_{\pi^-}$ are significantly positive at high $x_p$, indicating 
leading pion production.
The data give no information on the relative contributions of the three light
flavors, but are consistent with 0.27$D_{\rm p}$ at all $x_p$, and hence
with the notion
that leading pion effects in $u$ and $d$ jets are of similar strength to those
of leading baryons in all light quark jets.
Our measured $D_{K^-}$ are consistently positive for $x_p>0.15$, and well
above both 0.27$D_{\rm p}$ and $D_{\pi^-}$ at high $x_p$.
This indicates both production of leading charged kaons and greater 
production of leading $K^-$ in $s$-jets than in $\bar{u}$-jets.

The predictions of the three models are also shown in fig.~\ref{ndmc}.
For protons, the HERWIG prediction drops below zero for $0.05<x_p<0.25$, then
rises rapidly to unity at higher $x_p$; this structure is inconsistent with our
data.
The UCLA prediction also rises to unity at high $x_p$, becoming inconsistent
with the data for $x_p>$0.25.
The JETSET prediction is consistent with the data.
For the mesons, all models predict positive values of $D_{\pi^-}$ and $D_{K^-}$
even at very low $x_p$; they are all consistent with our data in that region,
but we are not able to resolve the difference from zero.
At high $x_p$, all three predictions for $D_{\pi^-}$ lie roughly 2$\sigma$
below the data,
and the measured $D_{K^-}$ favor the JETSET prediction slightly over
those of HERWIG and UCLA.

\section{Summary and Conclusions}

We have improved our measurements of the production of the charged hadron
species $\pi^\pm$, $K^\pm$ and p/$\bar{\rm p}$, as well as that of inclusive
stable charged particles, in hadronic $Z^0$ decays, taking advantage of the
sample of 400,000 decays recorded with the upgraded vertex detector to reduce
both the statistical and systematic errors substantially.
The SLD Cherenkov Ring Imaging Detector enabled the clean and efficient
identification of charged tracks over a wide momentum range,
yielding precise measurements of their total and differential production rates.
Our measurements in flavor-inclusive hadronic $Z^0$ decays are consistent with,
complementary to, and in some regions more precise than previous measurements.
Deficiencies in popular hadronization models have been confirmed.

The precision of the vertex detector allowed us to isolate very high-purity
light-, $c$- and $b$-tagged event samples, and to compare production
characteristics of the hadron species in light-, $c$- and $b$-flavor events.
Significant differences between flavors were found,
consistent with expectations based on the known properties of
$B$ and $D$ hadron production and decay.
The known problems with hadronization models were all present in the
light-flavor events, confirming that they are indeed in the hadronization itself
and not just artifacts of heavy hadron modelling.
Additional problems with the models were observed in $c$- and $b$-flavor events.

The shape of the $\xi = -\ln x_p$ distribution near its peak for each hadron
species in events of each flavor is consistent with the Gaussian form predicted
by MLLA QCD$+$LPHD.
For $c$- and $b$-flavor events, however, the Gaussian form cannot accommodate
the data over a wider range without the addition of very large distortion terms.
In light-flavor events, the Gaussian with small additional distortion terms is
able to describe the data over a wider range.
The peak positions $\xi^*$ for each hadron species in light-flavor events
are more consistent with a monotonic dependence on hadron mass than those in
flavor-inclusive events.
Our data are thus consistent with the predictions of MLLA QCD for the light
flavors, and indicate that the presence of heavy hadrons distorts the
observed spectra.

Using the large forward-backward asymmetry induced by the polarized SLC electron
beam to separate light quark from light antiquark hemispheres,
we have compared hadron and antihadron production in light quark jets.
A large excess of proton over antiproton production at high $x_p$ is direct
evidence for the production of leading baryons, i.e. baryons that carry the
quantum numbers of the quark that initiated the jet.
A similarly large excess of $K^-$ over $K^+$ production indicates not only
leading kaon production but also that leading kaons are produced more often
from initial $s$ quarks than from initial $u$ quarks.
A smaller excess of $\pi^-$ over $\pi^+$ production was also observed.
This is expected if leading pions account for a large fraction of high momentum
pion production but with equal $\pi^-$ rates from initial $d$ and $\bar{u}$
jets: $u\bar{u}$ and $d\bar{d}$ events then contribute to the observed
difference with opposite signs,
but since down-type quarks are produced more often and with a higher electroweak
asymmetry than up-type quarks, a net positive difference is observed.
These data provide unique and stringent tests of hadronization models.
All models tested were able to reproduce the pion and kaon data, though the
latter favor the JETSET model over the other two;  JETSET is also consistent
with the proton data, but the other two models predict values much higher than
the data for $x_p>$0.3.

\section*{Acknowledgements}

We thank the personnel of the SLAC accelerator department and the
technical
staffs of our collaborating institutions for their outstanding efforts
on our behalf.

\section*{$^{**}$List of Authors} 
%
%
%
\begin{center}
\def\iAOMORI{$^{(1)}$}
\def\iBRI{$^{(2)}$}
\def\iBRUN{$^{(3)}$}
\def\iBU{$^{(4)}$}
\def\iCOLO{$^{(5)}$}
\def\iCSU{$^{(6)}$}
\def\iFERR{$^{(7)}$}
\def\iFRAS{$^{(8)}$}
\def\iJHU{$^{(9)}$}
\def\iLBL{$^{(10)}$}
\def\iMASS{$^{(11)}$}
\def\iMISSI{$^{(12)}$}
\def\iMIT{$^{(13)}$}
\def\iMOSCOW{$^{(14)}$}
\def\iNAGO{$^{(15)}$}
\def\iOREG{$^{(16)}$}
\def\iOXF{$^{(17)}$}
\def\iPERU{$^{(18)}$}
\def\iQMU{$^{(19)}$}
\def\iRUTG{$^{(20)}$}
\def\iRAL{$^{(21)}$}
\def\iSLAC{$^{(22)}$}
\def\iSOONG{$^{(23)}$}
\def\iTENN{$^{(24)}$}
\def\iTOHO{$^{(25)}$}
\def\iUCSB{$^{(26)}$}
\def\iUCSC{$^{(27)}$}
\def\iVAND{$^{(28)}$}
\def\iWASH{$^{(29)}$}
\def\iWISC{$^{(30)}$}
\def\iYALE{$^{(31)}$}

  \baselineskip=.75\baselineskip
\mbox{Koya Abe\unskip,\iTOHO}
\mbox{Kenji Abe\unskip,\iNAGO}
\mbox{T. Abe\unskip,\iSLAC}
\mbox{I. Adam\unskip,\iSLAC}
\mbox{H. Akimoto\unskip,\iSLAC}
\mbox{D. Aston\unskip,\iSLAC}
\mbox{K.G. Baird\unskip,\iMASS}
\mbox{C. Baltay\unskip,\iYALE}
\mbox{H.R. Band\unskip,\iWISC}
\mbox{T.L. Barklow\unskip,\iSLAC}
\mbox{J.M. Bauer\unskip,\iMISSI}
\mbox{G. Bellodi\unskip,\iOXF}
\mbox{R. Berger\unskip,\iSLAC}
\mbox{G. Blaylock\unskip,\iMASS}
\mbox{J.R. Bogart\unskip,\iSLAC}
\mbox{G.R. Bower\unskip,\iSLAC}
\mbox{J.E. Brau\unskip,\iOREG}
\mbox{M. Breidenbach\unskip,\iSLAC}
\mbox{W.M. Bugg\unskip,\iTENN}
\mbox{D. Burke\unskip,\iSLAC}
\mbox{T.H. Burnett\unskip,\iWASH}
\mbox{P.N. Burrows\unskip,\iQMU}
\mbox{A. Calcaterra\unskip,\iFRAS}
\mbox{R. Cassell\unskip,\iSLAC}
\mbox{A. Chou\unskip,\iSLAC}
\mbox{H.O. Cohn\unskip,\iTENN}
\mbox{J.A. Coller\unskip,\iBU}
\mbox{M.R. Convery\unskip,\iSLAC}
\mbox{V. Cook\unskip,\iWASH}
\mbox{R.F. Cowan\unskip,\iMIT}
\mbox{G. Crawford\unskip,\iSLAC}
\mbox{C.J.S. Damerell\unskip,\iRAL}
\mbox{M. Daoudi\unskip,\iSLAC}
\mbox{S. Dasu\unskip,\iWISC}
\mbox{N. de Groot\unskip,\iBRI}
\mbox{R. de Sangro\unskip,\iFRAS}
\mbox{D.N. Dong\unskip,\iMIT}
\mbox{M. Doser\unskip,\iSLAC}
\mbox{R. Dubois\unskip,\iSLAC}
\mbox{I. Erofeeva\unskip,\iMOSCOW}
\mbox{V. Eschenburg\unskip,\iMISSI}
\mbox{E. Etzion\unskip,\iWISC}
\mbox{S. Fahey\unskip,\iCOLO}
\mbox{D. Falciai\unskip,\iFRAS}
\mbox{J.P. Fernandez\unskip,\iUCSC}
\mbox{K. Flood\unskip,\iMASS}
\mbox{R. Frey\unskip,\iOREG}
\mbox{E.L. Hart\unskip,\iTENN}
\mbox{K. Hasuko\unskip,\iTOHO}
\mbox{S.S. Hertzbach\unskip,\iMASS}
\mbox{M.E. Huffer\unskip,\iSLAC}
\mbox{X. Huynh\unskip,\iSLAC}
\mbox{M. Iwasaki\unskip,\iOREG}
\mbox{D.J. Jackson\unskip,\iRAL}
\mbox{P. Jacques\unskip,\iRUTG}
\mbox{J.A. Jaros\unskip,\iSLAC}
\mbox{Z.Y. Jiang\unskip,\iSLAC}
\mbox{A.S. Johnson\unskip,\iSLAC}
\mbox{J.R. Johnson\unskip,\iWISC}
\mbox{R. Kajikawa\unskip,\iNAGO}
\mbox{M. Kalelkar\unskip,\iRUTG}
\mbox{H.J. Kang\unskip,\iRUTG}
\mbox{R.R. Kofler\unskip,\iMASS}
\mbox{R.S. Kroeger\unskip,\iMISSI}
\mbox{M. Langston\unskip,\iOREG}
\mbox{D.W.G. Leith\unskip,\iSLAC}
\mbox{V. Lia\unskip,\iMIT}
\mbox{C. Lin\unskip,\iMASS}
\mbox{G. Mancinelli\unskip,\iRUTG}
\mbox{S. Manly\unskip,\iYALE}
\mbox{G. Mantovani\unskip,\iPERU}
\mbox{T.W. Markiewicz\unskip,\iSLAC}
\mbox{T. Maruyama\unskip,\iSLAC}
\mbox{A.K. McKemey\unskip,\iBRUN}
\mbox{R. Messner\unskip,\iSLAC}
\mbox{K.C. Moffeit\unskip,\iSLAC}
\mbox{T.B. Moore\unskip,\iYALE}
\mbox{M. Morii\unskip,\iSLAC}
\mbox{D. Muller\unskip,\iSLAC}
\mbox{V. Murzin\unskip,\iMOSCOW}
\mbox{S. Narita\unskip,\iTOHO}
\mbox{U. Nauenberg\unskip,\iCOLO}
\mbox{H. Neal\unskip,\iYALE}
\mbox{G. Nesom\unskip,\iOXF}
\mbox{N. Oishi\unskip,\iNAGO}
\mbox{D. Onoprienko\unskip,\iTENN}
\mbox{L.S. Osborne\unskip,\iMIT}
\mbox{R.S. Panvini\unskip,\iVAND}
\mbox{C.H. Park\unskip,\iSOONG}
\mbox{I. Peruzzi\unskip,\iFRAS}
\mbox{M. Piccolo\unskip,\iFRAS}
\mbox{L. Piemontese\unskip,\iFERR}
\mbox{R.J. Plano\unskip,\iRUTG}
\mbox{R. Prepost\unskip,\iWISC}
\mbox{C.Y. Prescott\unskip,\iSLAC}
\mbox{B.N. Ratcliff\unskip,\iSLAC}
\mbox{J. Reidy\unskip,\iMISSI}
\mbox{P.L. Reinertsen\unskip,\iUCSC}
\mbox{L.S. Rochester\unskip,\iSLAC}
\mbox{P.C. Rowson\unskip,\iSLAC}
\mbox{J.J. Russell\unskip,\iSLAC}
\mbox{O.H. Saxton\unskip,\iSLAC}
\mbox{T. Schalk\unskip,\iUCSC}
\mbox{B.A. Schumm\unskip,\iUCSC}
\mbox{J. Schwiening\unskip,\iSLAC}
\mbox{V.V. Serbo\unskip,\iSLAC}
\mbox{G. Shapiro\unskip,\iLBL}
\mbox{N.B. Sinev\unskip,\iOREG}
\mbox{J.A. Snyder\unskip,\iYALE}
\mbox{H. Staengle\unskip,\iCSU}
\mbox{A. Stahl\unskip,\iSLAC}
\mbox{P. Stamer\unskip,\iRUTG}
\mbox{H. Steiner\unskip,\iLBL}
\mbox{D. Su\unskip,\iSLAC}
\mbox{F. Suekane\unskip,\iTOHO}
\mbox{A. Sugiyama\unskip,\iNAGO}
\mbox{A. Suzuki\unskip,\iNAGO}
\mbox{M. Swartz\unskip,\iJHU}
\mbox{F.E. Taylor\unskip,\iMIT}
\mbox{J. Thom\unskip,\iSLAC}
\mbox{E. Torrence\unskip,\iMIT}
\mbox{T. Usher\unskip,\iSLAC}
\mbox{J. Va'vra\unskip,\iSLAC}
\mbox{R. Verdier\unskip,\iMIT}
\mbox{D.L. Wagner\unskip,\iCOLO}
\mbox{A.P. Waite\unskip,\iSLAC}
\mbox{S. Walston\unskip,\iOREG}
\mbox{A.W. Weidemann\unskip,\iTENN}
\mbox{E.R. Weiss\unskip,\iWASH}
\mbox{J.S. Whitaker\unskip,\iBU}
\mbox{S.H. Williams\unskip,\iSLAC}
\mbox{S. Willocq\unskip,\iMASS}
\mbox{R.J. Wilson\unskip,\iCSU}
\mbox{W.J. Wisniewski\unskip,\iSLAC}
\mbox{J.L. Wittlin\unskip,\iMASS}
\mbox{M. Woods\unskip,\iSLAC}
\mbox{T.R. Wright\unskip,\iWISC}
\mbox{R.K. Yamamoto\unskip,\iMIT}
\mbox{J. Yashima\unskip,\iTOHO}
\mbox{S.J. Yellin\unskip,\iUCSB}
\mbox{C.C. Young\unskip,\iSLAC}
\mbox{H. Yuta\unskip.\iAOMORI}

\it
  \vskip \baselineskip                   
  \baselineskip=.75\baselineskip   
\iAOMORI
  Aomori University, Aomori, 030 Japan, \break
\iBRI
  University of Bristol, Bristol, United Kingdom, \break
\iBRUN
  Brunel University, Uxbridge, Middlesex, UB8 3PH United Kingdom, \break
\iBU
  Boston University, Boston, Massachusetts 02215, \break
\iCOLO
  University of Colorado, Boulder, Colorado 80309, \break
\iCSU
  Colorado State University, Ft. Collins, Colorado 80523, \break
\iFERR
  INFN Sezione di Ferrara and Universita di Ferrara, I-44100 Ferrara, Italy,
\break
\iFRAS
  INFN Laboratori Nazionali di Frascati, I-00044 Frascati, Italy, \break
\iJHU
  Johns Hopkins University,  Baltimore, Maryland 21218-2686, \break
\iLBL
  Lawrence Berkeley Laboratory, University of California, Berkeley, California
94720, \break
\iMASS
  University of Massachusetts, Amherst, Massachusetts 01003, \break
\iMISSI
  University of Mississippi, University, Mississippi 38677, \break
\iMIT
  Massachusetts Institute of Technology, Cambridge, Massachusetts 02139, \break
\iMOSCOW
  Institute of Nuclear Physics, Moscow State University, 119899 Moscow, Russia,
\break
\iNAGO
  Nagoya University, Chikusa-ku, Nagoya, 464 Japan, \break
\iOREG
  University of Oregon, Eugene, Oregon 97403, \break
\iOXF
  Oxford University, Oxford, OX1 3RH, United Kingdom, \break
\iPERU
  INFN Sezione di Perugia and Universita di Perugia, I-06100 Perugia, Italy,
\break
\iQMU
  Queen Mary, University of London, London, E1 4NS United Kingdom,
\break
\iRUTG
  Rutgers University, Piscataway, New Jersey 08855, \break
\iRAL
  Rutherford Appleton Laboratory, Chilton, Didcot, Oxon OX11 0QX United Kingdom,
\break
\iSLAC
  Stanford Linear Accelerator Center, Stanford University, Stanford, California
94309, \break
\iSOONG
  Soongsil University, Seoul, Korea 156-743, \break
\iTENN
  University of Tennessee, Knoxville, Tennessee 37996, \break
\iTOHO
  Tohoku University, Sendai, 980 Japan, \break
\iUCSB
  University of California at Santa Barbara, Santa Barbara, California 93106,
\break
\iUCSC
  University of California at Santa Cruz, Santa Cruz, California 95064, \break
\iVAND
  Vanderbilt University, Nashville,Tennessee 37235, \break
\iWASH
  University of Washington, Seattle, Washington 98105, \break
\iWISC
  University of Wisconsin, Madison,Wisconsin 53706, \break
\iYALE
  Yale University, New Haven, Connecticut 06511. \break

\rm
%

\end{center}


\begin{thebibliography}{99}

\bibitem{ert}
See e.g. R.K. Ellis, D.A. Ross, A.E. Terrano, Nucl. Phys. {\bf B178} (1981) 421.

\bibitem{moretti}
S. Moretti, Phys. Lett. {\bf B420}~(1998)~367.

\bibitem{mlla} T.I.~Azimov, Y.L.~Dokshitzer, V.A.~Khoze and S.I.~Troyan, Z.
Phys. {\bf C27} (1985) 65.

\bibitem{nlla}
G. Marchesini and B.R. Webber, Nucl. Phys. {\bf B238} (1984) 1.

\bibitem{kkp}
See e.g.
B.A. Kniehl, G. Kramer and B. P\"{o}tter, Nucl. Phys. {\bf B582} (2000) 514.

\bibitem{saxon}
D.H. Saxon, {\it High Energy Electron-Positron Physics}, Eds. A. Ali and P.
S\"oding, World Scientific (1988), p. 539.

\bibitem{bohrer}
A. B\"ohrer, Phys. Rep. {\bf 291} (1997) 107.

\bibitem{pdg}
Particle Data Group, K. Hagiwara, et al., Phys. Rev. {\bf D66} (2002) 010001.

\bibitem{chmult}
SLD Collab., K. Abe, et al., Phys. Lett. {\bf B386} (1996) 475; \\
DELPHI Collab., P. Abreu, et al., Phys. Lett. {\bf B347} (1995) 447; \\
SLD Collab., K. Abe et al., Phys. Rev. Lett. {\bf 72} (1994) 3145.

\bibitem{chgxso} OPAL Collab., K. Akerstaff et al.,
E. Phys. J {\bf C7} (1999) 369.

\bibitem{lpprl} SLD Collab., K. Abe et al.,
Phys. Rev. Lett. {\bf 78} (1997) 3442.

\bibitem{bfp} SLD Collab., K. Abe et al.,
Phys. Rev. {\bf D59} (1999) 52001.

\bibitem{opallp} OPAL Collab., G. Abbiandi et al.,
Eur. Phys. J. {\bf C16} (2000) 407.

\bibitem{herwig}
G. Marchesini et al., Comp. Phys. Comm. {\bf 67}~(1992)~465.

\bibitem{jetset74}
T. Sj\"ostrand, Comp. Phys. Comm. {\bf 82}~(1994)~74.

\bibitem{ucla}
S. Chun and C. Buchanan, Phys. Rep. {\bf 292} (1998) 239.

\bibitem{sld} SLD Design Report, SLAC-Report 273 (1984).

\bibitem{cdc}
M.D. Hildreth et al., Nucl. Inst. Meth. {\bf A367} (1995) 111.

\bibitem{vxd3}
C. J. S. Damerell et al., Nucl. Inst. Meth. {\bf A400}~(1997)~287.

\bibitem{crid}
K. Abe et al., Nucl. Inst. Meth. {\bf A343} (1994) 74.

\bibitem{lac}
D. Axen et al., Nucl. Inst. Meth. {\bf A238} (1993) 472.

\bibitem{thrust}
S. Brandt et al., Phys. Lett. {\bf 12}~(1964)~57;\\
E. Farhi, Phys. Rev. Lett. {\bf 39}~(1977)~1587.

\bibitem{alr} SLD Collab., K. Abe et al.,
Phys. Rev. Lett. {\bf 73} (1994) 25.

\bibitem{hjkang}
H.J. Kang, Ph. D. Thesis, Rutgers University, SLAC-R-590 (2002).

\bibitem{davej}
D.J. Jackson, Nucl. Inst. Meth. {\bf A388} (1997) 247.

\bibitem{sldsim}
SLD Collaboration, K. Abe et al., Phys. Rev. Lett. {\bf 79} (1997) 590.

\bibitem{chgxsa} ALEPH Collab., R. Barate et al.,
Phys. Rept. {\bf 294} (1998) 1.

\bibitem{chgxsd} DELPHI Collab., P. Abreu et al.,
E. Phys. J {\bf C5} (1998) 585.

\bibitem {alept}
SLD Collab., K. Abe et al., Phys. Rev. Lett. {\bf 86}
(2001) 1162.

\bibitem{davea}
K. Abe et al., Nucl. Inst. and Meth. {\bf A371} (1996) 195.

\bibitem{tomp}
T.J. Pavel, Ph.D. Thesis, Stanford University, January 1997;
SLAC-Report-495.

\bibitem{opalfr}
OPAL Collab., P.D.~Acton et al., Z. Phys. {\bf C63}~(1994)~181.

\bibitem{bfrag}
See SLD Collab., K. Abe, et al., Phys. Rev. {\bf D65} (2002) 092006,
and references therein.


\end{thebibliography}
\end{document}